\documentclass[10pt]{article}

\usepackage{a4wide}
\usepackage{amssymb}
\usepackage{amsmath}
\usepackage{gensymb}
\usepackage{graphicx}
\usepackage{subfigure}
\usepackage[autostyle]{csquotes}
\usepackage{upgreek}
\usepackage{multirow}
\usepackage{hyperref}
\usepackage{xcolor}
\hypersetup{colorlinks, urlcolor = teal, linkcolor = teal, citecolor = teal}
\usepackage{algorithm}
\usepackage{algpseudocode}

\usepackage{authblk}

\begin{document}

\title{Mathematical Imaging Methods for Mitosis Analysis in Live-Cell Phase Contrast Microscopy}

\renewcommand{\thefootnote}{$\star$}
\author[a]{Joana Sarah Grah\footnote{Corresponding author, e-mail: \href{mailto:jg704@cam.ac.uk}{jg704@cam.ac.uk}}}
\author[b]{Jennifer Alison Harrington}
\author[b]{Siang Boon Koh\footnote{Present address: Massachusetts General Hospital Cancer Center, Harvard Medical School, Boston, MA 02114, USA}}
\author[b]{Jeremy Andrew Pike}
\author[b]{Alexander Schreiner\footnote{Present address: PerkinElmer, Schnackenburgallee 114, 22525 Hamburg, Germany}}
\author[c]{Martin Burger}
\author[a]{Carola-Bibiane Sch\"onlieb}
\author[b]{Stefanie Reichelt}

\affil[a]{University of Cambridge, Department of Applied Mathematics and Theoretical Physics, Centre for Mathematical Sciences, Wilberforce Road, Cambridge CB3 0WA, United Kingdom}
\affil[b]{University of Cambridge, Cancer Research UK Cambridge Institute, Li Ka Shing Centre, Robinson Way, Cambridge CB2 0RE, United Kingdom}
\affil[c]{Westf\"alische Wilhelms-Universit\"at M\"unster, Institute for Computational and Applied Mathematics, Einsteinstrasse 62, 48149 M\"unster, Germany}

\date{February 6, 2017}

\maketitle

\begin{abstract}
In this paper we propose a workflow to detect and track mitotic cells in time-lapse microscopy image sequences. In order to avoid the requirement for cell lines expressing fluorescent markers and the associated phototoxicity, phase contrast microscopy is often preferred over fluorescence microscopy in live-cell imaging. However, common specific image characteristics complicate image processing and impede use of standard methods. Nevertheless, automated analysis is desirable due to manual analysis being subjective, biased and extremely time-consuming for large data sets. Here, we present the following workflow based on mathematical imaging methods. In the first step, mitosis detection is performed by means of the circular Hough transform. The obtained circular contour subsequently serves as an initialisation for the tracking algorithm based on variational methods. It is sub-divided into two parts: in order to determine the beginning of the whole mitosis cycle, a backwards tracking procedure is performed. After that, the cell is tracked forwards in time until the end of mitosis. As a result, the average of mitosis duration and ratios of different cell fates (cell death, no division, division into two or more daughter cells) can be measured and statistics on cell morphologies can be obtained. All of the tools are featured in the user-friendly MATLAB$^{\textregistered}$ Graphical User Interface \emph{MitosisAnalyser}.
\end{abstract}

\section{Introduction}
\label{sec:introduction}

Mathematical image analysis techniques have recently become enormously important in biomedical research, which increasingly needs to rely on information obtained from images. Applications range from sparse sampling methods to enhance image acquisition through structure-preserving image reconstruction to automated analysis for objective interpretation of the data \cite{bioImageAnalysis}. In cancer research, observation of cell cultures in live-cell imaging experiments by means of sophisticated light microscopy is a key technique for quality assessment of anti-cancer drugs \cite{refJennyDrugs1,refJennyDrugs2}. In this context, analysis of the mitotic phase plays a crucial role. The balance between mitosis and apoptosis is normally carefully regulated, but many types of cancerous cells have evolved to allow uncontrolled cell division. Hence drugs targeting mitosis are used extensively during cancer chemotherapy. In order to evaluate the effects of a given drug on mitosis, it is desirable to measure average mitosis durations and distribution of possible outcomes such as regular division into two daughter cells, apoptosis, division into an abnormal number of daughter cells (one or $\geq$3) and no division at all \cite{reviewCellFates,fatesDrugs}.

Since performance of technical equipment such as microscopes and associated hardware is constantly improving and large amounts of data can be acquired in very short periods of time, automated image processing tools are frequently favoured over manual analysis, which is expensive and prone to error and bias. Generally, experiments might last several days and images are taken in a magnitude of minutes and from different positions. This leads to a sampling frequency of hundreds of images per sequence with an approximate size of 1000$^2$ pixels.

\subsection{Image Characteristics in Phase Contrast Microscopy}
\label{subsec:PCM}

In live-cell imaging experiments for anti-cancer drug assessment, the imaging modality plays a key role. Observation of cell cultures originating from specific cell lines under the microscope requires a particular setting ensuring that the cells do not die during image acquisition and that they behave as naturally as possible \cite{lightMicroscopy}. Here, phase contrast is often preferred to fluorescence microscopy because the latter requires labelling or transgenic expression of fluorescent markers, both causing phototoxicity and possibly changes of cell behaviour \cite{phototoxicity1,phototoxicity2,phototoxicity3}. As opposed to this, cells do not need to be stained for phase contrast microscopy. Moreover, phase shifts facilitate visualisation of even transparent specimens as opposed to highlighting of individual specific cellular components in fluorescence microscopy. We believe that one main advantage of our proposed framework is that it can be applied to data acquired with any standard phase-contrast microscope, which are prevalent in many laboratories and more widespread than for instance recently established quantitative phase imaging devices (e.g.\ Q-Phase by Tescan).

\begin{figure}[h]
\centering
\includegraphics[height=3cm]{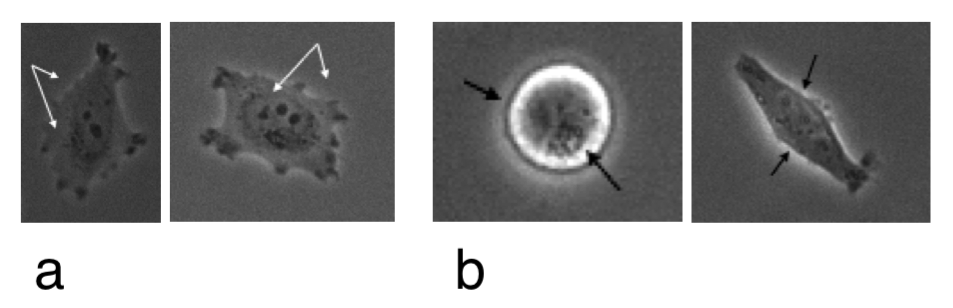}
\caption{Common image characteristics in phase contrast microscopy: shade-off effect (a) and halo effect (b) (HeLa DMSO control cells)}
\label{fig:effectsPCM}
\end{figure}

There are two common image characteristics occurring in phase contrast imaging (cf.\ Figure \ref{fig:effectsPCM}). Both visual effects highly impede image processing and standard algorithms are not applicable in a straightforward manner. The shade-off effect leads to similar intensities inside the cells and in the background. As a result, edges are only weakly pronounced and imaging methods such as segmentation relying on intensity gradient information (cf.\ Section \ref{subsubsec:GAC}) often fail. Moreover, region-based methods assuming that average intensities of object and background differ from one another (cf.\ Section \ref{subsubsec:ChanVese}) are not applicable either. Secondly, the halo effect is characterised by areas of high intensity surrounding cell membranes. The brightness levels increase significantly immediately before cells enter mitosis due to the fact that they round up, form a nearly spherically-shaped volume and therefore the amount of diffracted light increases. In addition, both effects prohibit application of basic image pre-processing tools like for example thresholding or histogram equalisation (cf.\ \cite{masterThesis}).

\subsection{Brief Literature Review}
\label{subsec:literature}

Over the past few years a lot of cell tracking frameworks have been established (cf.\ \cite{meijering2009tracking}) and some publications also feature mitosis detection. In \cite{ambuehl2012}, a two-step cell tracking algorithm for phase contrast images is presented, where the second step involves a level-set-based variational method. However, analysis of the mitotic phase is not included in this framework. Another tracking method based on extended mean-shift processes \cite{debeir2005} is able to incorporate cell divisions, but does not provide cell membrane segmentation. In \cite{huh2011} an automated mitosis detection algorithm based on a probabilistic model is presented, but it is not linked to cell tracking. A combined mitosis detection and tracking framework is established in \cite{thirusittampalam2013}, although cell outline segmentation is not included. Li et al. \cite{li2008} provide a comprehensive framework facilitating both tracking and lineage reconstruction of cells in phase contrast image sequences. Moreover, they are able to distinguish between mitotic and apoptotic events.

In addition, a number of commercial software packages for semi- or fully automated analysis of microscopy images exist, for example \textit{Volocity}, \textit{Columbus} (both PerkinElmer), \textit{Imaris} (Bitplane), \textit{ImageJ/Fiji} \cite{fiji} and \textit{Icy} \cite{icy} (also cf.\ \cite{BioImagingSoftwareTools}). The last two are open source platforms and the latter supports graphical protocols while the former incorporates a macro language, allowing for individualisation and extension of integrated tools. However, the majority of plugins and software packages are limited to analysis of fluorescence data.

A framework, which significantly influenced development of our methods and served as a basis for our tracking algorithm, was published in 2014 by M\"oller et al.\ \cite{SabinePaper}. It incorporates a MATLAB$^{\textregistered}$ Graphical User Interface that enables semi-automated tracking of cells in phase contrast microscopy time-series. The user has to manually segment the cells of interest in the first frame of the image sequence and can subsequently execute an automatic tracking procedure consisting of two rough and refined segmentation steps. In the following section, the required theoretical foundations of mathematical imaging methods are discussed, starting with the concept of the circular Hough transform and continuing with a review of segmentation and tracking methods leading to a more detailed description of the above-mentioned framework. For a more detailed discussion, we refer the interested reader to \cite{masterThesis} and the references therein.

\section{Mathematical Background}
\label{sec:MB}

\subsection{The Circular Hough Transform}
\label{subsec:CHT}

The Hough transform is a method for automated straight line recognition in images patented by Paul Hough in 1962 \cite{Hough}. It was further developed and generalised by Duda and Hart in 1972 \cite{DudaHart}. More specifically, they extended the Hough transform to different types of parametrised curves and in particular, they applied it to circle detection.

The common strategy is to transform points lying on straight line segments or curves in the underlying image into a parameter space. Its dimension depends on the number of variables required in order to parametrise the sought-after curve. For the parametric representation of a circle, which can be written as
\begin{equation}
r^2 = (x - c_1)^2 + (y - c_2)^2,
\label{eq:circle}
\end{equation}
the radius $r$ as well as two centre coordinates $(c_1,c_2)$ are required. Hence, the corresponding parameter space is three-dimensional. Each point $(x,y)$ in the original image satisfying the above equation for fixed $r$, $c_1$ and $c_2$ coincides with a cone in the parameter space. Then, edge points of circular objects in the original image correspond to intersecting cones and from detecting those intersections in the parameter space one can again gather circles in the image space.

\begin{figure}[h]
\centering
\includegraphics[height=3cm]{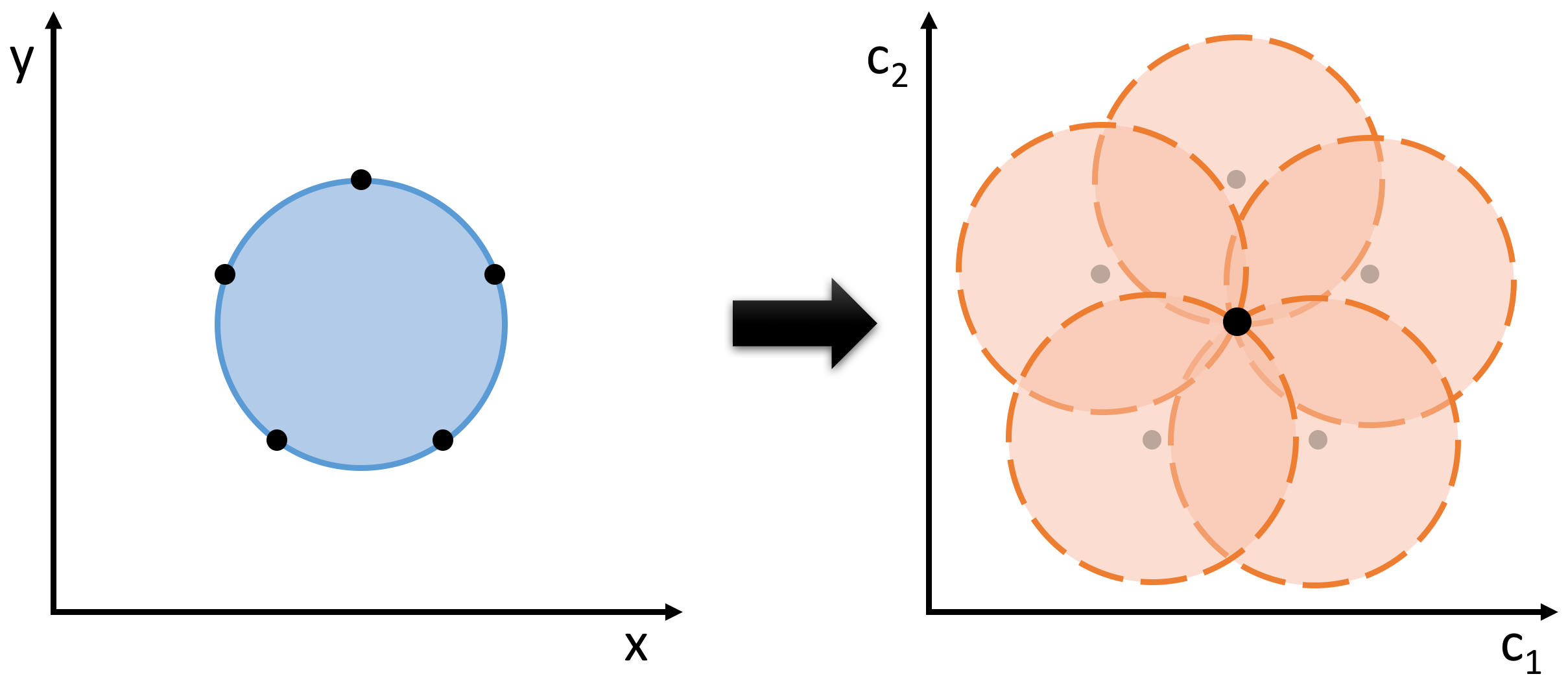}
\caption{The circular Hough transform}
\label{fig:cht}
\end{figure}

For simplification, we fix the radius and consider the two-dimensional case in Figure \ref{fig:cht}. On the left, we have the image space, i.e.\ the $x$-$y$-plane, and a circle in light blue with five arbitrary points located on its edge highlighted in dark blue. All points fulfil equation \eqref{eq:circle} for fixed centre coordinates $(c_1,c_2)$. On the other hand, fixing those specific values for $c_1$ and $c_2$ in the parameter space, i.e.\ $c_1$-$c_2$-plane, on the right, and keeping $x$ and $y$ in \eqref{eq:circle} arbitrary, leads to the dashed orange circles, where the corresponding edge points are drawn in grey for orientation. All of the orange circles intersect in one point, which exactly corresponds to the circle centre in the original image. Hence, from intersections in the parameter space one can reference back to circular objects in the image space.

A discussion on how the circular Hough transform is embedded and implemented in \emph{MitosisAnalyser} can be found in Section \ref{subsec:MD}.

\subsection{Image Segmentation and Tracking}
\label{subsec:tracking}

In the following, we would like to introduce variational methods (cf.\ e.g.\ \cite{AubertKornprobst,ChanShen2005}) for imaging problems. The main aim is minimisation of an energy functional modelling certain assumptions on the given data and being defined as
\begin{equation}
\label{eq:energyFunctional}
E(\phi) = D(\phi,\psi) + \alpha R(\phi).
\end{equation}
It is dependent on the solution $\phi$, which represents the processed image to be obtained, and shall be minimised with respect to $\phi$. The given image to be processed is denoted by $\psi$. The functions $\phi$ and $\psi$ map from the rectangular image domain $\Omega \subset \mathbb{R}^2$ to $\Sigma \subset \mathbb{R}^d$ containing colour ($d=3$) or greyscale ($d=1$) intensity values. In the case of 8-bit phase contrast microscopy images, $d=1$ and $\Sigma = \{0,\dots,255\}$, where 0 and 255 correspond to black and white, respectively.

The first part $D$ on the right-hand side of \eqref{eq:energyFunctional} ensures data fidelity between $\phi$ and $\psi$, i.e. the solution $\phi$ should be reasonably close to the original input data $\psi$. This can be obtained by minimising a norm measuring the distance between $\psi$ and $\phi$, where the choice of norm naturally depends on the given problem. The regulariser $R$ in \eqref{eq:energyFunctional} incorporates a-priori knowledge about the function $\phi$. For example, $\phi$ could be constrained to be sufficiently smooth in a particular sense. The parameter $\alpha$ is weighting the two different terms and thereby defines which one is considered to be more important. Energy functionals can also consist of multiple data terms and regularisers. Eventually, a solution that minimises the energy functional \eqref{eq:energyFunctional} attains a small value of $D$ assuring high fidelity to the original data, of course depending on the weighting. Similarly, a solution which has a small value of $R$ can be interpreted as having a high coincidence with the incorporated prior assumptions.

Here, we focus on image segmentation. The goal is to divide a given image into associated parts, e.g.\ object(s) and background. This can be done by finding either the objects themselves or the corresponding edges, which is then respectively called region-based and edge-based segmentation. However, those two tasks are very closely related and even coincide in the majority of cases. Tracking can be viewed as an extension of image segmentation because it describes the process of segmenting a sequence of images or video. The goal of object or edge identification remains the same, but the time-dependence is an additional challenge.

Below, we briefly discuss the level-set method and afterwards present two well-established segmentation models incorporating the former. Furthermore, we recap the methods in \cite{SabinePaper} building upon the above and laying the foundations for our proposed tracking framework.

\subsubsection{The Level-Set Method}
\label{subsubsec:LevelSet}

In 1988 the level-set method was introduced by Osher and Sethian \cite{OsherSethian1988}. The key idea is to describe motion of a front by means of a time-dependent partial differential equation. In variational segmentation methods, energy minimisation corresponds to propagation of such a front towards object boundaries. In two dimensions, a segmentation curve $c$ is modelled as the zero-level of a three-dimensional level-set function $\phi$. Two benefits are straightforward numerical implementation without need of parametrisation and implicit modelling of topological changes of the curve. The level-set evolution equation can be written as
\begin{equation*}
\frac{\partial \phi}{\partial t} = F \cdot \vert \nabla \phi \vert
\end{equation*}
with curvature-dependent speed of movement $F$ and suitable initial and boundary conditions.

\begin{figure}[h]
\centering
\includegraphics[height=3cm]{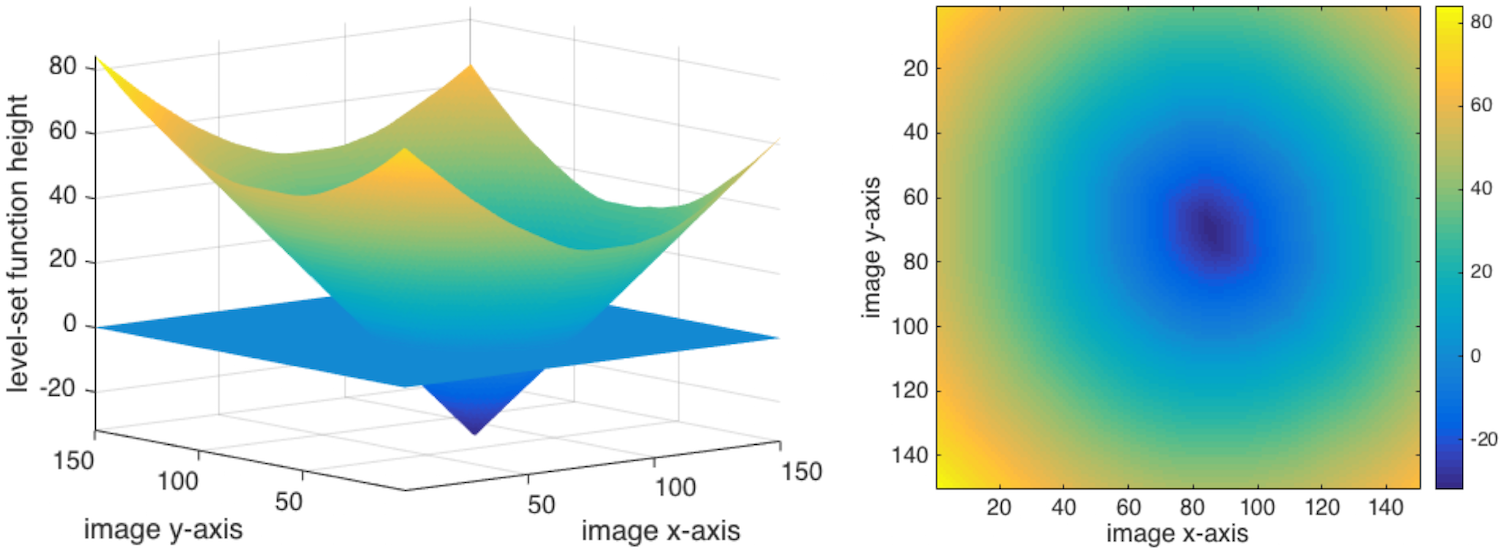}
\caption{Level-set function}
\label{fig:levelSet}
\end{figure}

For implementation, the level-set function $\phi$ is assigned negative values inside and positive values outside of the curve $c$,
\begin{equation}
\label{eq:phi}
\phi(t,x)
\begin{cases}
<0, & \text{if}\ x\ \text{is inside of}\ c,\\
=0, & \text{if}\ x\ \text{lies on}\ c,\\
>0, & \text{if}\ x\ \text{is outside of}\ c,
\end{cases}
\end{equation}
commonly chosen to be the signed Euclidean distances (cf.\ Figure \ref{fig:levelSet}).

\subsubsection{Geodesic Active Contours}
\label{subsubsec:GAC}

Active contours or \enquote{snakes} have been developed and extended for decades \cite{snakes1993caselles,ACballoons,snakes1995kichenassamy,geodesicac2000,snakes1988kass} and belong to the class of edge-based segmentation methods. As the name suggests, the goal is to move segmentation contours towards image edges and stop at boundaries of objects to be segmented (e.g.\ by using the level-set method described above). Geodesic active contours constitute a specific type of active contours methods and have been introduced by Caselles, Kimmel and Sapiro in 1997 \cite{geodesicac1997}. The level-set formulation reads
\begin{equation}
\frac{\partial \phi}{\partial t} = \underbrace{\nabla \cdot \left(g \frac{\nabla\phi}{\vert\nabla\phi\vert} \right)}_{F}\ \cdot\ \vert \nabla \phi \vert\
\label{eq:GAC}
\end{equation}
with appropriate initial and boundary conditions and $g$ is an edge-detector function typically depending on the gradient magnitude of a smoothed version of a given image $\psi$. A frequently used function is
\begin{equation}
\label{eq:EdgeDetectorFunctiong}
g = \frac{1}{1 + \vert \nabla (G_{\sigma}*\psi(x)) \vert^2}
\end{equation}
with $G_{\sigma}$ being a Gaussian kernel with standard deviation $\sigma$. The function $g$ is close to zero at edges, where the gradient magnitude is high, and close or equal to one in homogeneous image regions, where the gradient magnitude is nearly or equal to zero. Hence, the segmentation curve, i.e.\ the zero-level of $\phi$, propagates towards edges defined by $g$ and once the edges are reached, evolution is stopped. In the specific case of $g=1$, \eqref{eq:GAC} coincides with mean curvature motion.

Geodesic active contours are a well-suited method of choice for segmentation if image edges are strongly pronounced or can otherwise be appropriately identified by a suitable function $g$.

\subsubsection{Active Contours without Edges}
\label{subsubsec:ChanVese}

As the name suggests, the renowned model developed by Chan and Vese in 2001 \cite{ChanVese} is a region-based segmentation method and in contrast to the model presented in \ref{subsubsec:GAC}, edge information is not taken into account. It is rather based on the assumption that the underlying image can be partitioned into two regions of approximately piecewise-constant intensities. In the level-set formulation the variational energy functional reads
\begin{align}
\label{eq:ChanVese}
\begin{split}
E(\phi,c_1,c_2) =&\ \lambda_1 \int\limits_{\Omega} \left( \psi(x) - c_1 \right)^2 \left( 1 - H(\phi(x)) \right)\,dx + \lambda_2 \int\limits_{\Omega} \left( \psi(x) - c_2 \right)^2 H(\phi(x))\,dx\\
&+ \mu \int\limits_{\Omega} \left|\nabla H(\phi(x))\right|\,dx + \nu \int\limits_{\Omega} \left(1-H(\phi(x))\right)\,dx,
\end{split}
\end{align}
which is to be minimised with respect to $\phi$ as well as $c_1$ and $c_2$. Recalling \eqref{eq:phi}, we define the Heaviside function $H$ as
\begin{equation}
\label{eq:Heaviside}
H(\phi)=
\begin{cases}
0, & \text{if}\ \phi \leq 0,\\
1, & \text{if}\ \phi > 0,
\end{cases}
\end{equation}
indicating the sign of the level-set function and therefore the position relative to the segmentation curve. 

In \eqref{eq:ChanVese} the structure in \eqref{eq:energyFunctional} is resembled. The first two data terms enforce a partition into two regions with intensities $c_1$ inside and $c_2$ outside of the segmentation contour described by the zero-level-set. The third and fourth terms are contour length and area regularisers, respectively.

The optimal $c_1$ and $c_2$ can be directly calculated while keeping $\phi$ fixed:
\begin{equation*}
c_1 = \frac{\int_{\Omega} \psi(x) \left( 1-H(\phi(x)) \right)\,dx}{\int_{\Omega} \left( 1-H(\phi(x)) \right)\,dx}, \quad c_2 = \frac{\int_{\Omega} \psi(x) H(\phi(x))\,dx}{\int_{\Omega} H(\phi(x))\,dx}.
\end{equation*}
In order to find the optimal $\phi$ and hence the sought-after segmentation contour, the Euler-Lagrange equation defined as $\frac{\partial \phi}{\partial t} = -\frac{\partial E}{\partial \phi} = 0$ needs to be calculated, which leads to the evolution equation
\begin{align}
\label{eq:ChanVeseSystem}
\frac{\partial \phi}{\partial t} = \delta_{\varepsilon}(\phi) \left( \lambda_1 \left( \psi - c_1 \right)^2 - \lambda_2 \left( \psi - c_2 \right)^2 + \mu\ \nabla \cdot \left( \frac{\nabla\phi}{|\nabla\phi|} \right) + \nu \right),
\end{align}
where $\delta_{\varepsilon}$ is the following regularised version of the Dirac delta function:
\begin{equation*}
\label{eq:DeltaRegularisation}
\delta_{\varepsilon}(\phi)=\frac{\varepsilon}{\pi}\left(\varepsilon^2+\phi^2\right).
\end{equation*}
Equation \eqref{eq:ChanVeseSystem} can be numerically solved with a gradient descent method.

This model is very advantageous for segmenting noisy images with weakly pronounced or blurry edges as well as objects and clustering structures of different intensities in comparison to the background.

\subsubsection{Tracking Framework by M\"oller et al.}
\label{subsubsec:Sabine}

The cell tracking framework developed in \cite{SabinePaper} is sub-divided into two steps. First, a rough segmentation based on the model in Section \ref{subsubsec:ChanVese} is performed. The associated energy functional reads
\begin{align}
\label{eq:SabineEnergyFunctional}
\begin{split}
E(\phi,c_1,c_2) =&\ \lambda_1 \int_{\Omega} \left( | v | - c_1 \right)^2 \left( 1 - H(\phi(x)) \right)\,dx + \lambda_2 \int_{\Omega} \left( | v | - c_2 \right)^2 H(\phi(x))\,dx\\
&+ \mu \int_{\Omega} | \nabla H(\phi(x)) |\,dx + \nu \left( \int_{\Omega} \left( 1 - H(\phi(x)) \right)\,dx - V_{\text{old}} \right)^2.
\end{split}
\end{align}
In contrast to \eqref{eq:ChanVese}, the area or volume regularisation term weighted by $\nu$ is altered such that the current volume shall be close to the previous volume $V_{\text{old}}$. Moreover, the data terms weighted by $\lambda_1$ and $\lambda_2$ incorporate the normal velocity image $\vert v \vert$ instead of the image intensity function $\psi$:
\begin{equation}
\vert v \vert = \frac{\left\vert \frac{\partial}{\partial t} \psi \right\vert}{\vert \nabla \psi \vert_{\varepsilon}},
\label{eq:normVel}
\end{equation}
where the expression in the denominator is a regularisation of the gradient magnitude defined as $\vert \nabla \psi \vert_{\varepsilon} = \sqrt{(\partial_{x_1} \psi)^2 + (\partial_{x_2} \psi)^2 + \varepsilon^2}$ for small $\varepsilon$. The novelty here is that in contrast to only considering the image intensity both spatial and temporal information is used in order to perform the region-based segmentation. Indeed, cells are expected to move between subsequent frames. In addition, the gradient magnitude shall be increased in comparison to background regions. Therefore the incorporation of both temporal and spatial derivative provides a better indicator of cellular interiors.

In a second step, a refinement is performed using the geodesic active contours equation \eqref{eq:GAC}. The edge-detector function is customised and mainly uses information obtained by the Laplacian of Gaussian of the underlying image. In addition, topology is preserved throughout the segmentation by using the simple points scheme \cite{topology1994bertrand,topology2003han,topology1989kong} and in order to reduce computational costs this is combined with a narrow band method \cite{NarrowBandMethod}, which we inherit in our framework as well.

\section{\emph{MitosisAnalyser} Framework}
\label{sec:MiA}

In the following we present our proposed workflow designed in order to facilitate mitosis analysis in live-cell phase contrast imaging experiments. We specifically focused on applicability and usability while providing a comprehensive tool that needs minimal user interaction and parameter tuning. The MATLAB\textsuperscript{\textregistered} Graphical User Interface \emph{MitosisAnalyser} (The corresponding code is available at \url{github.com/JoanaGrah/MitosisAnalyser}.) provides a user-friendly application, which involves sets of pre-determined parameters for different cell lines and has been designed for non-experts in mathematical imaging.

\begin{figure}[h]
\centering
\includegraphics[height=4cm]{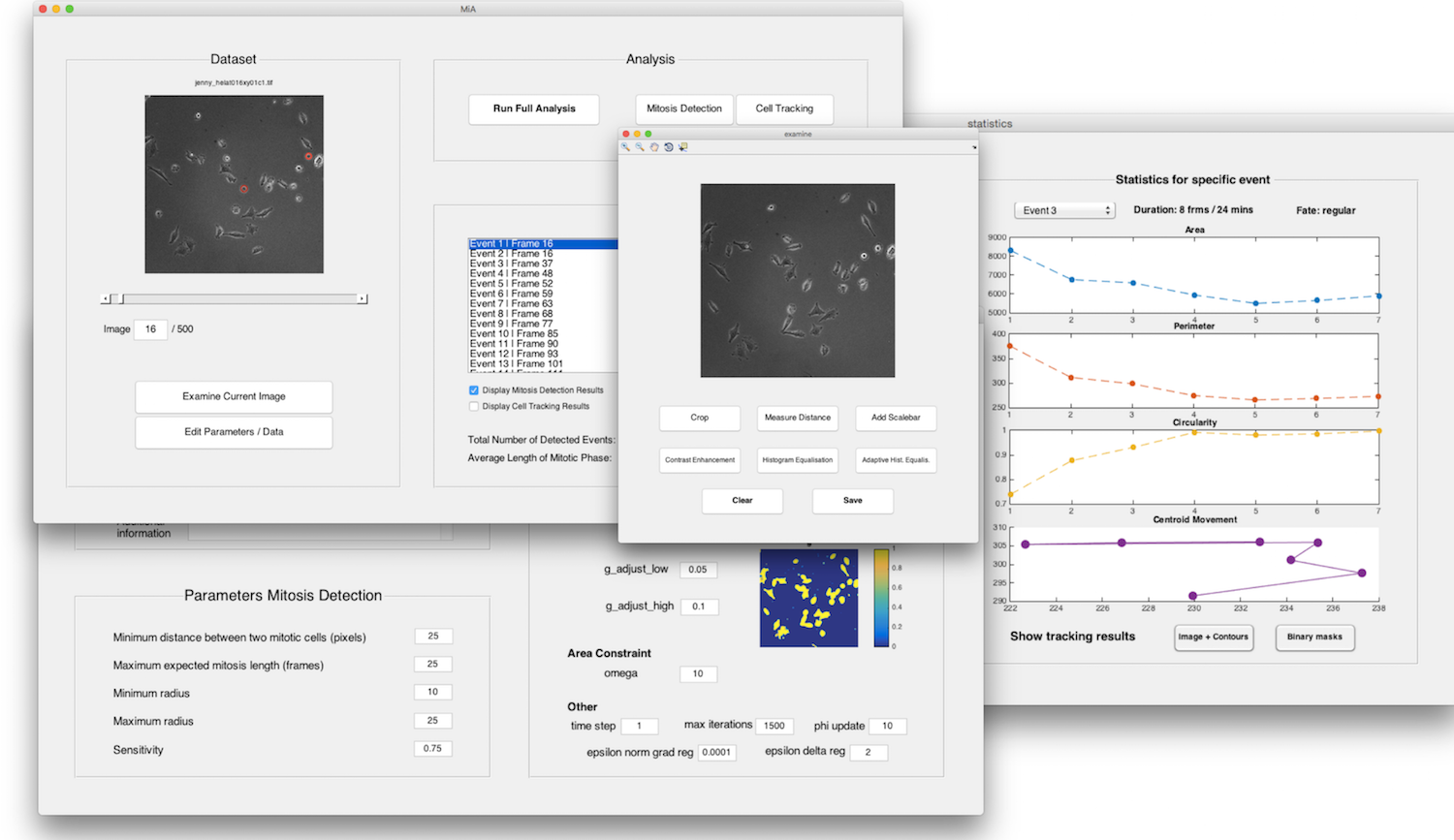}
\caption{\emph{MitosisAnalyser} MATLAB\textsuperscript{\textregistered} GUI}
\label{fig:mitosisanalyser}
\end{figure}

In Figure \ref{fig:mitosisanalyser} the main application window is displayed on the top left. The entire image sequence at hand can be inspected and after analysis, contours are overlaid for immediate visualisation. Moreover, images can be examined and pre-processed by means of a few basic tools (centre), although the latter did not turn out to be necessary for our types of data. Parameters for both mitosis detection and tracking can be reviewed, adapted and permanently saved for different cell lines in another separate window (bottom left). Mitosis detection can be run separately and produces intermediate results, where all detected cells can be reviewed and parameters can be adjusted as required. Consecutively, running the cell tracking algorithm results in an estimate of average mitosis duration and provides the possibility to survey further statistics (right).

\begin{figure}[h]
\centering
\includegraphics[height=5cm]{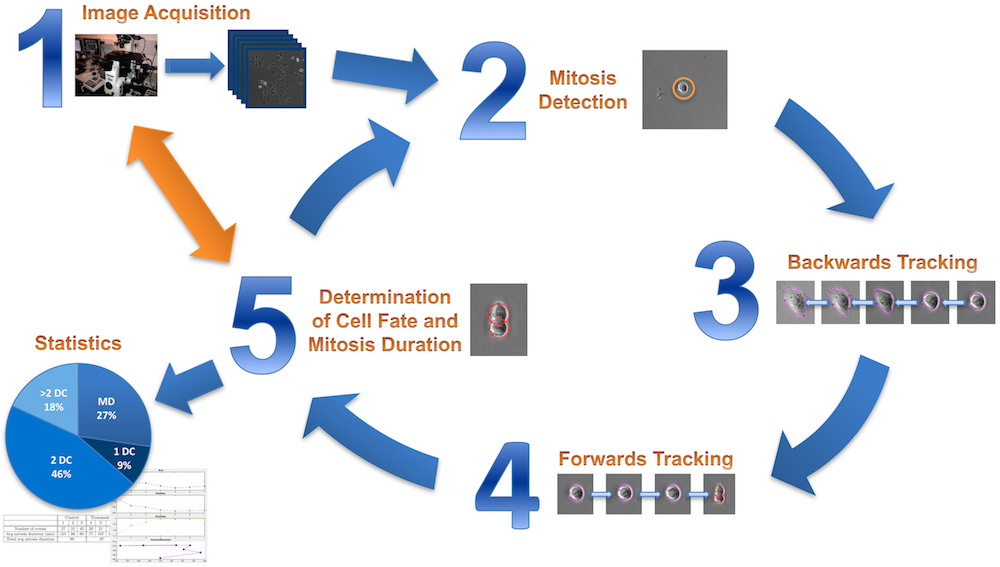}
\caption{Summary of \emph{MitosisAnalyser} framework}
\label{fig:miaSummary}
\end{figure}

Figure \ref{fig:miaSummary} summarises the entire workflow from image acquisition to evaluation of results. First, live-cell imaging experiments are conducted using light microscopy resulting in 2D greyscale image sequences. Next, mitosis detection is performed. For each detected cell, steps 3-5 are repeated. Starting at the point in time where the cell is most circular, the circle-shaped contour serves as an initialisation for the segmentation. The tracking is then performed backwards in time, using slightly extended contours from previous frames as initialisations. As soon as cell morphology changes, i.e. area increases and circularity decreases below a predetermined threshold, the algorithm stops and marks the point in time at hand as start of mitosis. Subsequently, again starting from the detected mitotic cell, tracking is identically performed forwards in time until the cell fate can be determined. As already mentioned in Section \ref{sec:introduction}, different cases need to be distinguished from one another: regular, abnormal and no division as well as apoptosis. The final step comprises derivation of statistics on mitosis duration and cell fate distribution as well as evaluation and interpretation thereof.

The double arrow connecting steps 1 and 5 indicates what is intended to be subject of future research. Ideally, image analysis shall be performed in on-line time during image acquisition and intermediate results shall be passed on to inform and influence microscopy software. Consequently, this may in turn lead to enhancement of image processing. Recently established concepts of bilevel optimisation and parameter learning for variational imaging models (cf. \cite{bilevelLearningCCRSV,bilevelLearningKP}) might supplement our framework.

\subsection{Mitosis Detection}
\label{subsec:MD}

In order to implement the circular Hough transform (CHT) described in Section \ref{subsec:CHT}, both image and parameter space need to be discretised. The former is naturally already represented as a pixel grid or matrix of grey values. The latter needs to be artificially discretised by binning values for $r$, $c_1$ and $c_2$ and the resulting representation is called accumulator array. Once the CHT is performed for all image pixels, the goal is to find peaks in the accumulator array referring to circular objects.

There are several options in order to speed up the algorithm, but we will only briefly discuss two of them. First, it is common to perform edge detection on the image before applying the CHT, since pixels lying on a circle very likely correspond to edge pixels. An edge map can for instance be calculated by thresholding the gradient magnitude image in order to obtain a binary image. Then, only edge pixels are considered in the following steps. Furthermore, it is possible to reduce the accumulator array to two dimensions using the so-called phase-coding method. The idea is using complex values in the accumulator array with the radius information encoded in the phase of the array entries. Both enhancements are included in the built-in MATLAB\textsuperscript{\textregistered} function \texttt{imfindcircles}.

\begin{figure}[h]
\centering
\includegraphics[height=2.25cm]{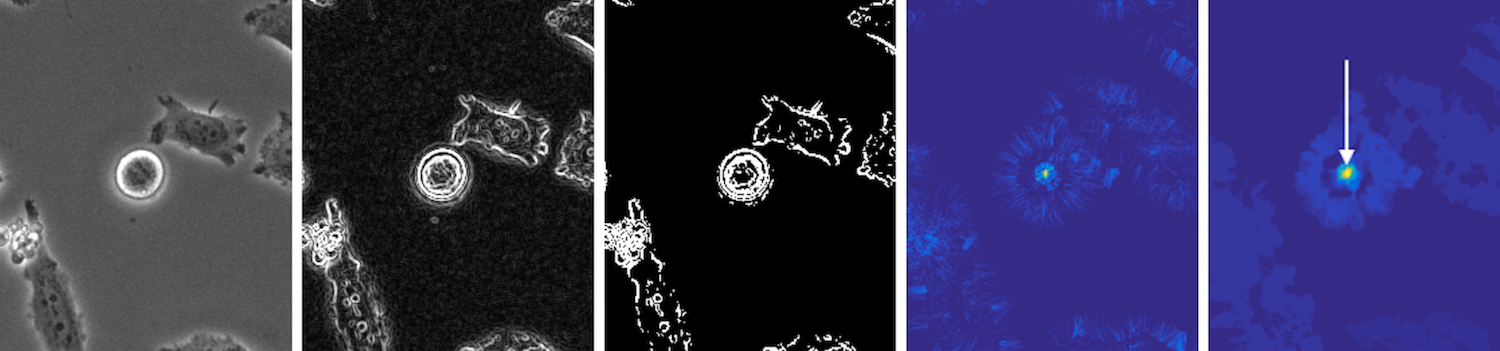}
\caption{Finding circles by means of the CHT. From left to right: Original greyscale image, gradient image, edge pixels, accumulator matrix, transformed matrix}
\label{fig:imfindcircles}
\end{figure}

The mitosis detection algorithm implemented into \emph{MitosisAnalyser} uses this function in order to perform the CHT and search for circular objects in the given image sequences. Figure \ref{fig:imfindcircles} visualises the different steps from calculation of the gradient image, to identification of edge pixels, to computation of the accumulator matrix and transformation thereof by filtering and thresholding, to detection of maxima.

This method turned out to be very robust and two main advantages are that circles of different sizes can be found and even not perfectly circularly shaped or overlapping objects can be detected. At the beginning of analysis, the CHT is applied in every image of the given image sequence in order to detect nearly circularly shaped mitotic cells. Afterwards, the circles are sorted by significance, which is related to the value of the detected peak in the corresponding accumulator array. The most significant ones are picked while simultaneously ensuring that identical cells are neither detected multiple times in the same frame nor in consecutive frames. The complete procedure is outlined in Supplementary Algorithm \ref{alg:md}.

\subsection{Cell Tracking}
\label{subsec:cellTracking}

We have already introduced variational segmentation methods in general as well as three models our framework is based on in more detail in Section \ref{subsec:tracking}. Here, we would like to state the cell tracking model we developed starting from the one presented in Section \ref{subsubsec:Sabine}. The energy functional reads:
\begin{align}
\label{eq:miaEnergyFunctional}
\begin{split}
E(\phi,c_1,c_2) =&\ \lambda_1 \int_{\Omega} \left( | v | - c_1 \right)^2 \left( 1 - H(\phi(x)) \right)\,dx + \lambda_2 \int_{\Omega} \left( | v | - c_2 \right)^2 \left( H(\phi(x)) \right)\,dx\\
&+ \mu \int_{\Omega} \left|\nabla H(\phi(x))\right|\,dx + \nu \int_{\Omega} g(\psi(x)) \left|\nabla H(\phi(x))\right|\,dx\\
&- \omega \frac{1}{2} \max \left\{ \int_{\Omega} \left( 1 - H(\phi(x)) \right)\,dx - t_{\text{area}}, 0 \right\}^2,
\end{split}
\end{align}
with $\vert v \vert$ and $H$ defined as in \eqref{eq:normVel} and \eqref{eq:Heaviside}, respectively.

The two terms weighted by $\lambda_1$ and $\lambda_2$ are identical to the ones in \eqref{eq:SabineEnergyFunctional}. Instead of having two separate segmentation steps as in \cite{SabinePaper}, we integrate the edge-based term weighted by $\nu$ into our energy functional. However, using a common edge-detector function based on the image gradient like the one in \eqref{eq:EdgeDetectorFunctiong} was not suitable for our purposes. We noticed that the gradient magnitude image contains rather weakly pronounced image edges, which motivated us to search for a better indicator of the cells' interiors. We realised that the cells are very inhomogeneous in contrast to the background and consequently, we decided to base the edge-detector function on the local standard deviation of grey values in a 3$\times$3-neighbourhood around each pixel. Additionally smoothing the underlying image with a standard Gaussian filter and rescaling intensity values leads to an edge-detector function, which is able to indicate main edges and attract the segmentation contour towards them.

Furthermore, we add a standard length regularisation term weighted by $\mu$. We complement our energy functional with an area regularisation term that incorporates a-priori information about the approximate cell area and prevents contours from becoming too small or too large. This penalty method facilitates incorporation of a constraint in the energy functional and in this case the area shall not fall below the threshold $t_{\text{area}}$.

\begin{figure}[h]
\centering
\includegraphics[height=7cm]{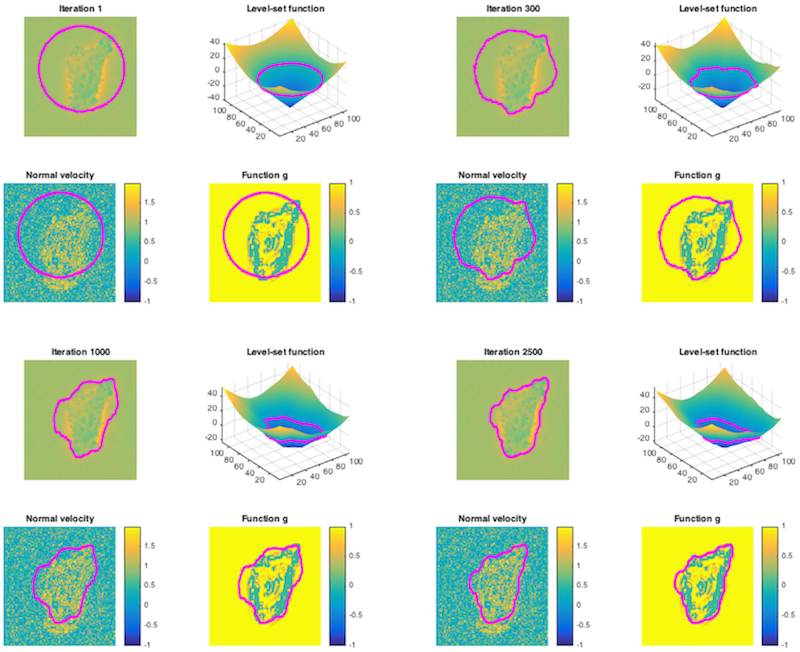}
\caption{Level-set evolution from initialisation to final iteration}
\label{fig:levelSetEvolution}
\end{figure}

Optimal parameters $c_1$ and $c_2$ can be calculated directly. We numerically minimise \eqref{eq:miaEnergyFunctional} with respect to the level-set function $\phi$ by using a gradient descent method (cf.\ \ref{subsubsec:ChanVese}). The third term weighted by $\mu$ is discretised using a combination of forwards, backwards and central finite differences as proposed in \cite{ChanVese}. We obtain the most stable numerical results by applying central finite differences to all operators contained in the fourth term weighted by $\nu$. In Figure \ref{fig:levelSetEvolution} we visualise level-set evolution throughout the optimisation procedure.

In order to give an overview of the backwards and forwards tracking algorithms incorporated in the mitosis analysis framework, we state the procedures in Supplementary Algorithm \ref{alg:bt} and \ref{alg:ft}. Together with the mitosis detection step they form the foundation of the routines included in \emph{MitosisAnalyser}.

\section{Material and Methods}
\label{sec:MM}

The \emph{MitosisAnalyser} framework is tested in three experimental settings with MIA PaCa-2 cells, HeLa Aur A cells and T24 cells. Below, a description of cell lines and chemicals is followed by details on image acquisition and standard pre-processing.

\subsection{Cell Lines and Chemicals}

The FUCCI(\textbf{F}luorescent \textbf{U}biquitination-based \textbf{C}ell \textbf{C}ycle \textbf{I}ndicator \cite{fucci})-expressing MIA PaCa-2 cell line was generated using the FastFUCCI reporter system and has previously been characterised and described \cite{siang2016,publicationSiang}. Cells were cultured in phenol red-free Dulbecco's modified Eagle's medium (DMEM) supplemented with 10\% foetal calf serum (FBS).

T24 cells were acquired from CLS. The T24 cells were cultured in DMEM/F12 (1:1) medium supplemented with 5\% FBS.

HeLa Aur A cells, HeLa cells modified to over-express aurora kinase A, were generated by Dr Jennifer Harrington with Dr David Perera at the Medical Research Council Cancer Unit, Cambridge, using the Flp-In T-REx system from Invitrogen as described before \cite{refJennyHela}. The parental HeLa LacZeo/TO line, and pOG44 and pcDNA5/FRT/TO plasmids were kindly provided by Professor Stephen Taylor, University of Manchester.  The parental line grows under selection with 50 $\upmu$g/ml Zeocin\texttrademark (InvivoGen) and 4 $\upmu$g/ml Blasticidin (Invitrogen). HeLa Aur A cells were cultured in DMEM supplemented with 10\% FBS and 4 $\upmu$g/ml blasticidin (Invitrogen) and 200 $\upmu$g/ml hygromycin (Sigma Aldrich). Transgene expression was achieved by treatment with 1 $\upmu$g/ml doxycycline (Sigma Aldrich).

In all experiments, all cells were grown at 37$\degree$C and 5\% CO$_2$ up to a maximum of 20 passages and for fewer than 6 months following resuscitation. They were also verified to be mycoplasma-free using the Mycoprobe$^{\textregistered}$ Mycoplasma Detection Kit (R\&D Systems). Paclitaxel (Tocris Bioscience), MLN8237 (Stratech Scientific) and Docetaxel (Sigma Aldrich) were dissolved in dimethylsulphoxide (DMSO, Sigma) in aliquots of 30mM, kept at -20$\degree$C and used within 3 months. Final DMSO concentrations were kept constant in each experiment ($\leq$0.2\%).

\subsection{Acquisition and Processing of Live-Cell Time-Lapse Sequences}

Cells were seeded in $\upmu$-Slide glass bottom dish (ibidi) and were kept in a humidified chamber under cell culture conditions (37$\degree$C, 5\% CO$_2$). For experiments with T24 and HeLa Aur A cells they were cultured for 24 hours before being treated with drugs or DMSO control. They were then imaged for up to 72 hours. Images were taken from three to five fields of view per condition, every 5 minutes, using a Nikon Eclipse TE2000-E microscope with a 20X (NA 0.45) long-working distance air objective, equipped with a sCMOS Andor Neo camera acquiring $2048\times2048$ images, which have been binned by a factor of two. Red and green fluorescence of the FUCCI-expressing cells were captured using a pE-300white CoolLED source of light filtered by Nikon FITC B-2E/C and TRITC G-2E/C filter cubes, respectively. For processing, an equalisation of intensities over time was applied to each channel, followed by a shading correction and a background subtraction, using the NIS-Elements software (Nikon).

\section{Results and Discussion}
\label{sec:results}

In this section we present and discuss results obtained by applying \emph{MitosisAnalyser} to the aforementioned experimental live-cell imaging data. A list of parameters we chose can be found in Supplementary Table \ref{tab:parameters}. For each cell line, we established a unique set of parameters. Nevertheless, the individual values are in reasonable ranges and do not differ significantly from one another. We did not follow a specific parameter choice rule, but rather tested various combinations and manually picked the best performing ones.

\subsection{MIA PaCa-2 Cells}
\label{subsec:FUCCI}

In a multi-modal experiment with FUCCI-expressing MIA PaCa-2 cells, both phase contrast images and fluorescence data were acquired. The latter consist of two channels with red and green intensities corresponding to CDT1 and Geminin signals, respectively. In this case we do use fluorescence microscopy imaging data as well, but we would like to stress that this analysis would not have been possible without the mitosis detection and tracking performed on the phase contrast data. As before, mitotic cells are detected using the circular Hough transform applied to the phase contrast images. Cell tracking is performed on the phase contrast images as well, but in addition, information provided by the green fluorescent data channel is used. More specifically, stopping criteria for both backwards and forwards tracking are based on green fluorescent intensity distributions indicating different stages of the cell cycle, which can be observed and is described in more detail in Supplementary Figure \ref{fig:plots_fluorescence_intensities}.

\begin{figure}[h]
\centering
\includegraphics[width=\textwidth]{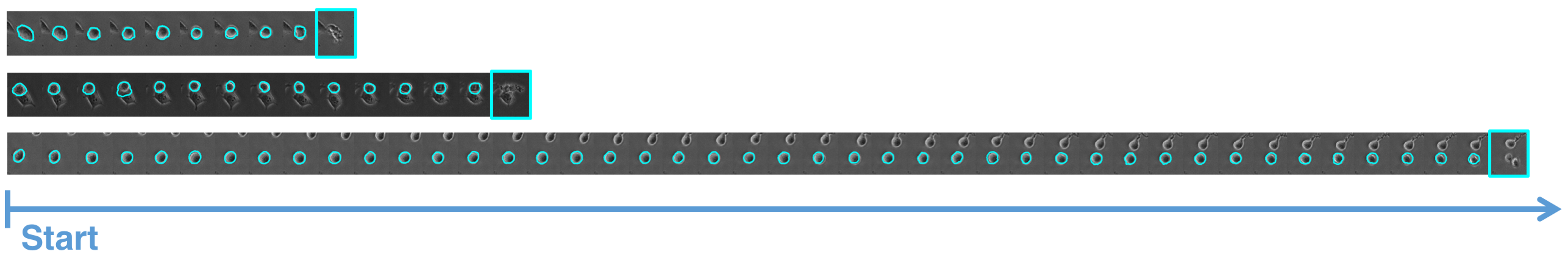}
\caption{Three examples of mitotic events detected for FUCCI MIA PaCa-2 \enquote{DMSO control}, \enquote{treatment with 3nM paclitaxel} and \enquote{treatment with 30nM paclitaxel} data (from top to bottom)}
\label{fig:FUCCItimeline}
\end{figure}

\begin{table}[h]
\centering
\begin{tabular}{|c||c|c|c||c|c|c||c|c|c|}
\hline & \multicolumn{3}{|c||}{DMSO control} & \multicolumn{3}{|c||}{3nM paclitaxel} & \multicolumn{3}{|c|}{30nM paclitaxel}\\
\hline & Pos 1 & Pos 2 & Pos 3 & Pos 4 & Pos 5 & Pos 6 & Pos 7 & Pos 8 & Pos 9\\
\hline
\hline Events & 14 & 11 & 13 & 12 & 8 & 19 & 10 & 13 & 35\\
\hline AMD & 51 & 41 & 60 & 52 & 88 & 94 & 146 & 104 & 112\\
\hline Total AMD & \multicolumn{3}{|c||}{51} & \multicolumn{3}{|c||}{78} & \multicolumn{3}{|c|}{121}\\
\hline
\end{tabular}
\caption{Average Mitosis Durations (AMD) for MIA PaCa-2 cell line in minutes}
\label{tab:FUCCI_AMD}
\end{table}

The whole data set consists of nine imaging positions, where three at a time correspond to DMSO control, treatment with 3nM paclitaxel and treatment with 30nM paclitaxel. Figure \ref{fig:FUCCItimeline} visualises exemplary courses of the mitotic phase, which could be measured by means of our proposed workflow. Table \ref{tab:FUCCI_AMD} presents estimated average mitosis durations for the three different classes of data. Indeed, the average duration of 51 minutes for the control is consistent with that obtained from manual scoring (cf.\ \cite{publicationSiang}, Figure S3D). Moreover, we can observe a dose-dependent increase in mitotic duration for the two treatments, which was anticipated, since paclitaxel leads to mitotic arrest.

\subsection{HeLa Cells}
\label{subsec:HeLa}

In the following we discuss results achieved by applying \emph{MitosisAnalyser} to sequences of phase contrast microscopy images showing HeLa Aur A cells. In addition to DMSO control data, cells have been treated with 25nM MLN8237 (MLN), 0.75nM paclitaxel (P), 30nM paclitaxel (P) and with a combination of 25nM MLN8237 and 0.75nM paclitaxel (combined).

\begin{figure}[h]
\centering
\includegraphics[height=4.5cm]{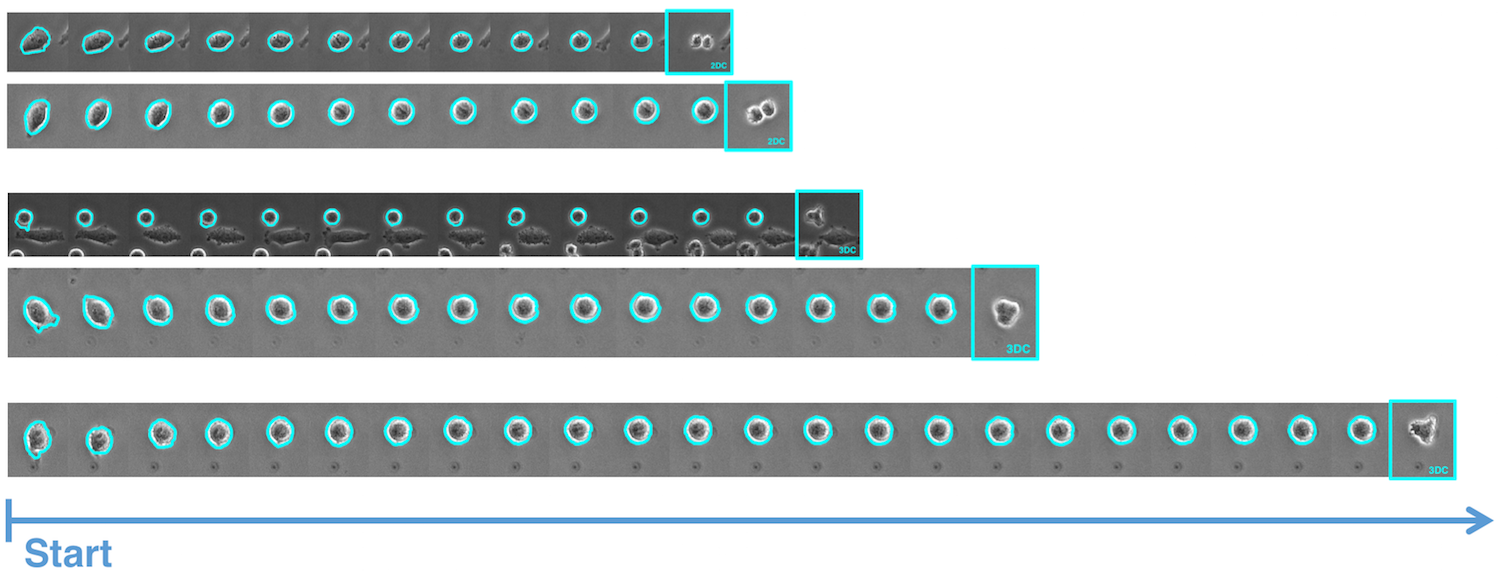}
\caption{Five examples of mitotic events detected for HeLa Aur A \enquote{DMSO control} (one each in row one and two), \enquote{treatment with 25nM MLN8237} (one each in row three and four), and \enquote{combined treatment with 25nM MLN8237 and 0.75nM paclitaxel} (bottom row) data}
\label{fig:HeLatimeline}
\end{figure}

Figure \ref{fig:HeLatimeline} shows exemplary results for detected and tracked mitotic events, where DMSO control cells divide regularly into two daughter cells. Particular treatments are expected to enhance multipolar mitosis and indeed our framework was able to depict the three daughter cells in each of the three examples (bottom rows) presented. In addition, mitosis duration is extended, as anticipated, for treated cells and specifically for the combined treatment. The segmentation of the cell membranes seems to work well by visual inspection, even in the case of touching neighbouring cells.

\begin{table}[h]
\centering
\begin{tabular}{|c||c||c||c||c||c|}
\hline & DMSO & 25nM MLN & 0.75nM P & 30nM P & Combined\\
\hline
\hline Events & 44 & 75 & 10 & 35 & 43\\
\hline AMD & 58 & 73 & 68 & 116 & 105\\
\hline
\end{tabular}
\caption{Average Mitosis Durations (AMD) for HeLa cell line in minutes}
\label{tab:HeLa_AMD}
\end{table}

Table \ref{tab:HeLa_AMD} summarises average mitosis durations that have been estimated for the different treatments. Again, the results are according to our expectations, i.e.\ mitosis durations for treated cells are extended in comparison to DMSO control.

\subsection{T24 Cells}
\label{subsec:T24}

\begin{figure}[h]
\centering
\includegraphics[height=3cm]{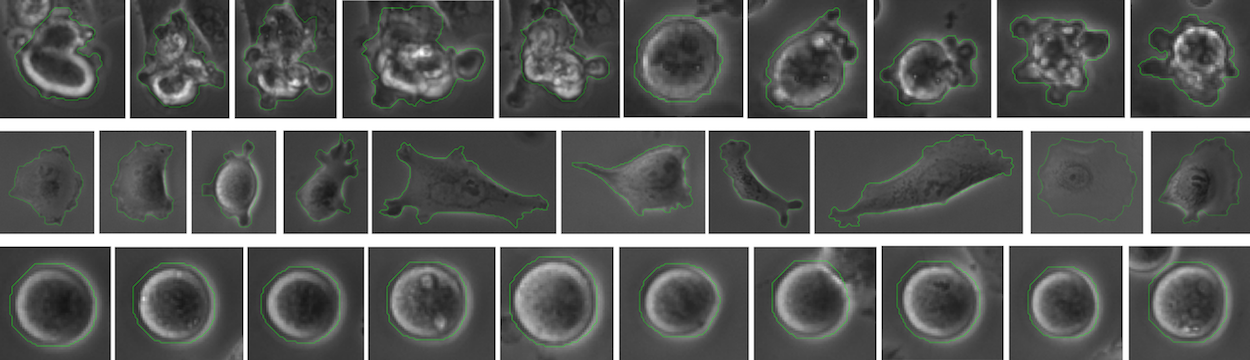}
\caption{Three manually segmented classes of T24 cells: apoptotic (top row), flat/normal (middle row) and mitotic (bottom row)}
\label{fig:classes}
\end{figure}

For this data set we wanted to focus on cell fate determination and in order to distinguish between different fates in the T24 cell data set we combine the \emph{MitosisAnalyser} framework with basic classification techniques. In particular, we manually segmented three different classes of cells: mitotic and apoptotic ones as well as cells in their normal state outside of the mitotic cell cycle phase (see Figure \ref{fig:classes}).

\begin{figure}[h]
\centering
\includegraphics[height=8cm]{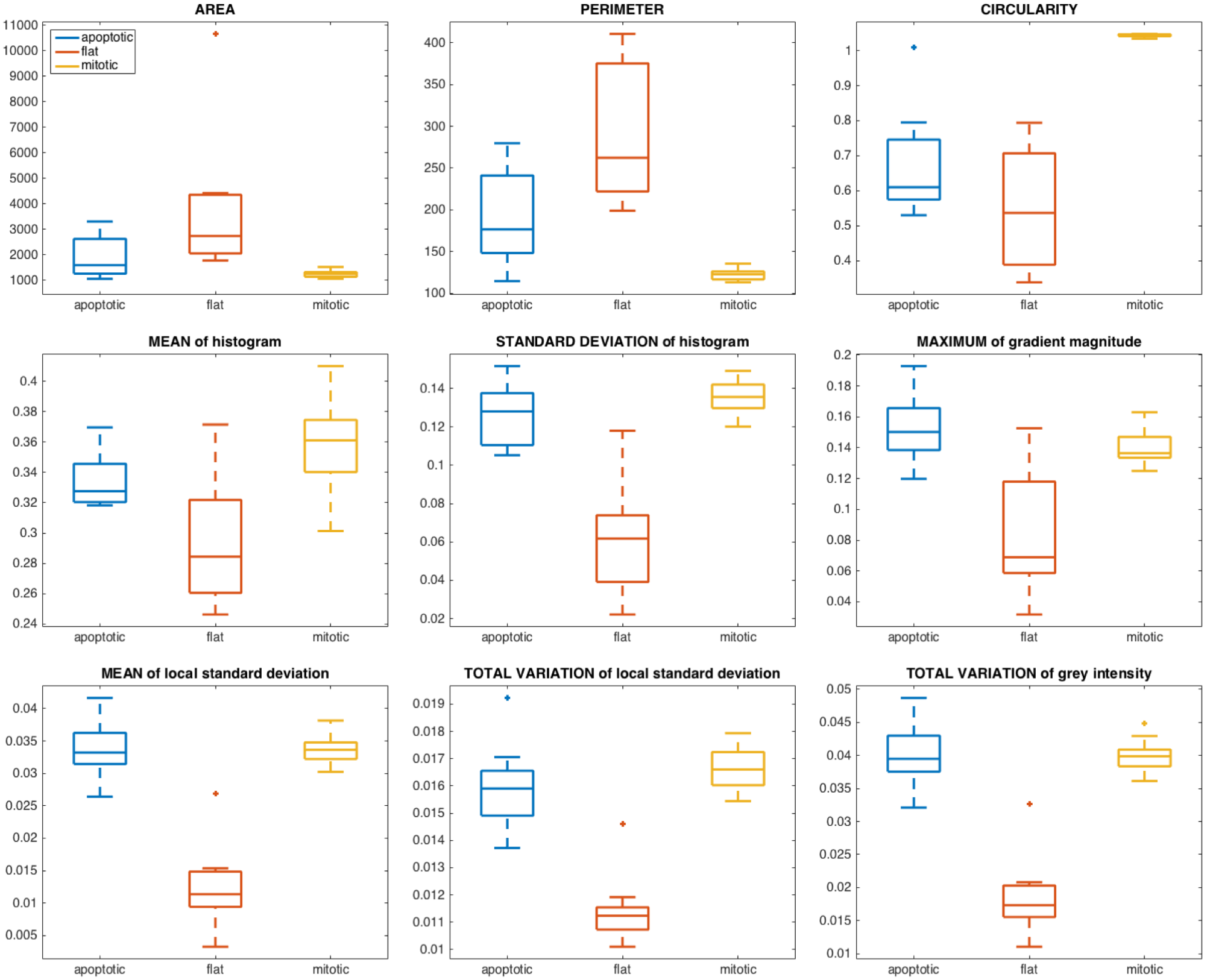}
\caption{Key features for cell type classification}
\label{fig:boxplots}
\end{figure}

In Figure \ref{fig:boxplots} we show boxplots of nine features based on morphology as well as intensity values we use for classification. Those include area, perimeter and circularity. Furthermore, we calculate both mean and standard deviation of the histogram. In addition, we consider the maximum of the gradient magnitude, the mean as well as the total variation of the local standard deviation and the total variation of the grey values. One can clearly observe that cells in mitosis have much higher circularity than in any other state. Flat cells differ significantly from the other two classes with respect to features based on intensity values.

In order to train a classifier solely based on those few features we used the MATLAB$^{\textregistered}$ Machine Learning Toolbox and its accompanying Classification Learner App. We chose a nearest-neighbour classifier with the number of neighbours set to 1 using Euclidean distances and equal distance weights, which yielded a classification accuracy of 93.3\% (cf.\ Supplementary Figure \ref{fig:plotsClassification}).

Pie charts for T24 cell fate distributions for different drug treatments as preliminary results can be found in Supplementary Figure \ref{fig:pieChartsT24}, although integration of classification techniques will be subject of more extensive future research.

\subsection{Validation}
\label{subsec:validation}

In order to validate performance of the segmentation, we compare results obtained with \emph{MitosisAnalyser} with blind manual segmentation. For that purpose, we choose two different error measures: The Jaccard Similarity Coefficient (JSC) \cite{JSC} and the Modified Hausdorff Distance (MHD) \cite{MHD}, which we are going to define in the following.

Let A and M be the sets of pixels included in the automated and manual segmentation mask, respectively. The JSC is defined as
\begin{equation*}
\text{JSC}(A,M) = \frac{|A \cap M|}{|A \cup M|},
\end{equation*}
where $A \cap M$ denotes the intersection of sets A and M, which contains pixels that are elements of both A and M. The union of sets A and M, denoted by $A \cup M$, contains pixels that are elements of A or M, i.e. elements either only of A or only of M or of $A \cap M$.
The MHD is a generalisation of the Hausdorff distance, which is commonly used to measure distance between shapes. It is defined as
\begin{equation*}
\text{MHD}(A,M) = \max \left\{ \frac{1}{\vert A \vert} \sum\limits_{a \in A} d(a,M), \frac{1}{\vert M \vert} \sum\limits_{m \in M} d(m,A) \right\},
\end{equation*}
where $d(a,M) = \min_{m \in M} \Vert a - m \Vert$ with Euclidean distance $\Vert \cdot \Vert$.

\begin{figure}[h]
\centering
\includegraphics[height=3cm]{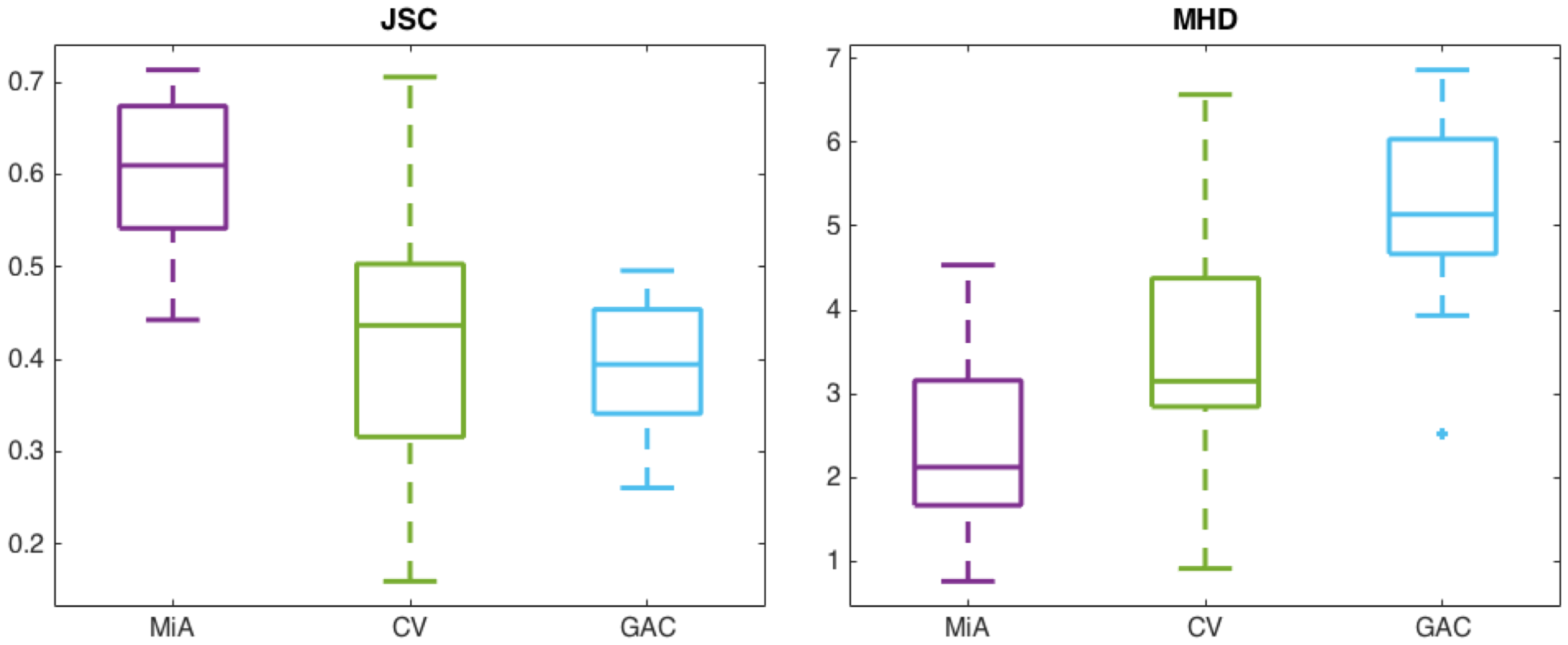}
\caption{Boxplots showing JSC (left) and MHD (right) measures for segmentation of apoptotic cell images by \emph{MitosisAnalyser} (MiA), the model by Chan and Vese (CV) and geodesic active contours (GAC) in comparison with manual segmentation}
\label{fig:boxplotsSegmentation}
\end{figure}

\begin{figure}[h]
\centering
\includegraphics[height=2cm]{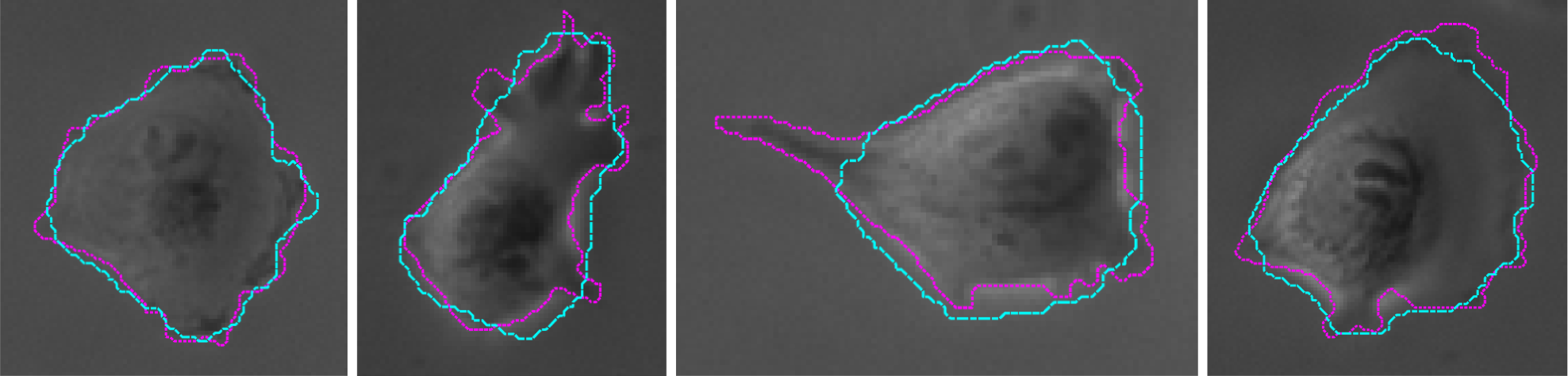}
\caption{Exemplary segmentations for flat cells in phase contrast images: Manual segmentation (magenta) is compared to performance of \emph{MitosisAnalyser} (cyan). The average JSC and MHD values for the four images are 0.8377 and 0.3648, respectively.}
\label{fig:segmentationFlatCells}
\end{figure}

The JSC assumes values between 0 and 1 and the closer it is to 1 the better is the segmentation quality. The MHD on the other hand is equal to 0 if two shapes coincide and the larger the number, the farther they differ from each other. In Figure \ref{fig:boxplotsSegmentation} and Supplementary Table \ref{tab:qualityMeasures} we can observe that on average, \emph{MitosisAnalyser} performs better than the standard Chan-Vese method (cf.\ Section \ref{subsubsec:ChanVese}) and Geodesic Active Contours based on the gradient magnitude (cf.\ Section \ref{subsubsec:GAC}) (both performed using the MATLAB \texttt{imageSegmenter} application) compared to manual segmentation of ten apoptotic T24 cell images (cf.\ Figure \ref{fig:classes}, top row). Moreover, Figure \ref{fig:segmentationFlatCells} shows successful segmentation of flat T24 cells affected by the shade-off effect in phase contrast microscopy images using \emph{MitosisAnalyser}, where both the method by Chan and Vese and geodesic active contours failed.

\subsection{Conclusions}
\label{subsec:conclusions}

We have used concepts of mathematical imaging including the circular Hough transform and variational tracking methods in order to develop a framework that aims at detecting mitotic events and segmenting cells in phase contrast microscopy images, whilst overcoming the difficulties associated with those images. Originating from the models presented in Section \ref{sec:MB}, we developed a customised workflow for mitosis analysis in live-cell imaging experiments performed in cancer research and discussed results we obtained by applying our methods to different cell line data.

\section{Acknowledgements}

JSG acknowledges support by the NIHR Cambridge Biomedical Research Centre and would like to thank Hendrik Dirks, Fjedor Gaede \cite{FjedorBT} and Jonas Geiping \cite{JonasBT} for fruitful discussions in the context of a practical course at WWU M\"unster in 2014 and significant speed-up and GPU implementation of earlier versions of the code. JSG and MB would like to thank Michael M\"oller for providing the basic tracking code and acknowledge support by ERC via Grant EU FP 7 - ERC Consolidator Grant 615216 LifeInverse. MB acknowledges further support by the German Science Foundation DFG via Cells-in-Motion Cluster of Excellence. CBS acknowledges support from the EPSRC grant Nr.~EP/M00483X/1, from the Leverhulme grant \enquote{Breaking the non-convexity barrier}, from the EPSRC Centre for Mathematical And Statistical Analysis Of Multimodal Clinical Imaging grant Nr. EP/N014588/1, and the Cantab Capital Institute for the Mathematics of Information. JAH, SBK, JAP, AS and SR were funded by Cancer Research UK, The University of Cambridge and Hutchison Whampoa Ltd. SBK also received funding from Pancreatic Cancer UK.

\bibliographystyle{unsrt}
\bibliography{arXiv}

\appendix
\setcounter{figure}{0}
\setcounter{table}{0}
\floatname{algorithm}{Supplementary Algorithm}
\renewcommand{\figurename}{Supplementary Figure}
\renewcommand{\tablename}{Supplementary Table}
\renewcommand\thefigure{\thesection\arabic{figure}}
\renewcommand\thetable{\thesection\arabic{table}}
\renewcommand\thealgorithm{\thesection\arabic{algorithm}}
\renewcommand\theHfigure{\thesection\thefigure} 
\renewcommand\theHtable{\thesection\thetable}

\section{Supplementary Data}

\begin{figure}[h!]
\centering
\subfigure[Position 1 - DMSO control]{\includegraphics[height=6.5cm]{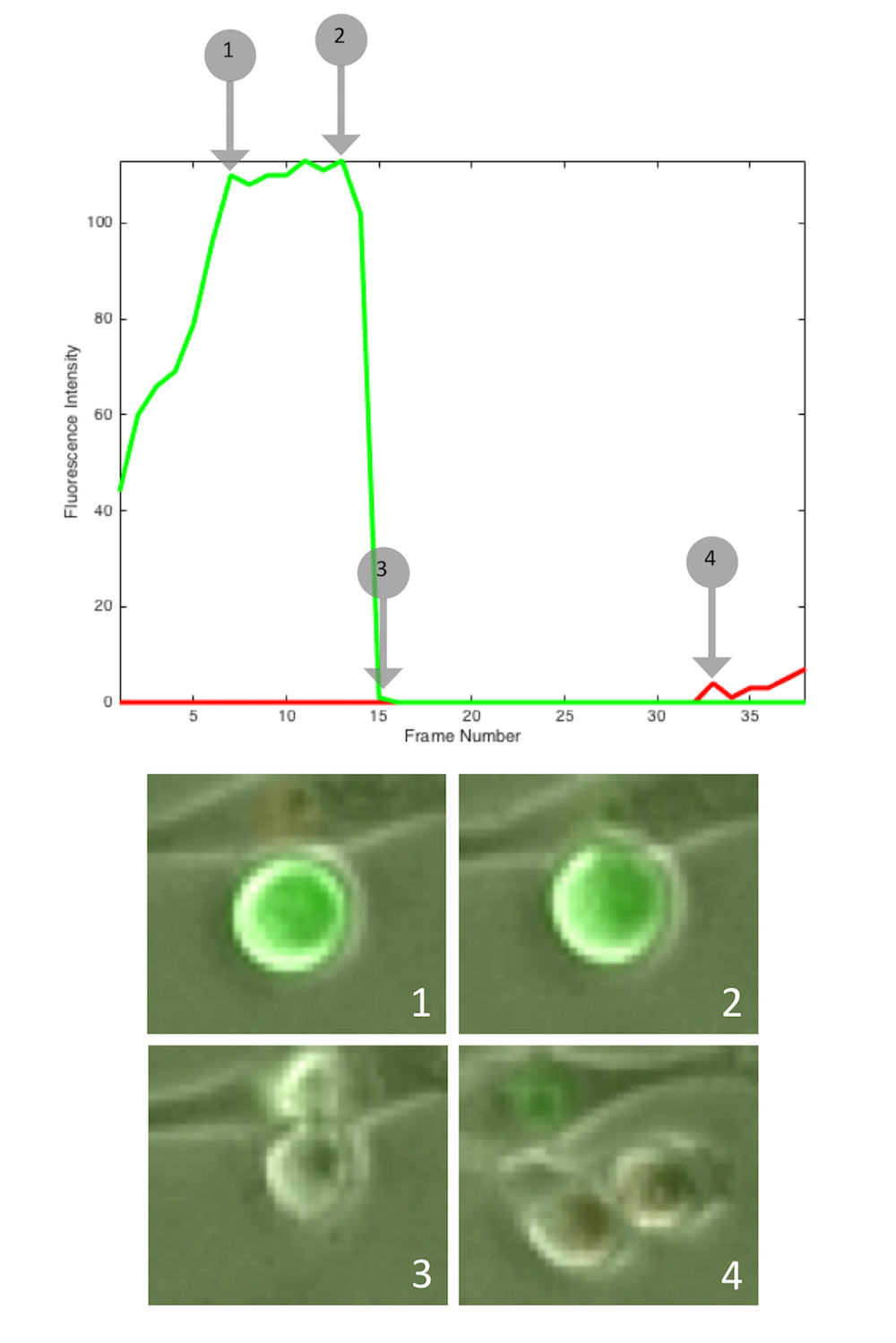}}
\subfigure[Position 2 - DMSO control]{\includegraphics[height=6.5cm]{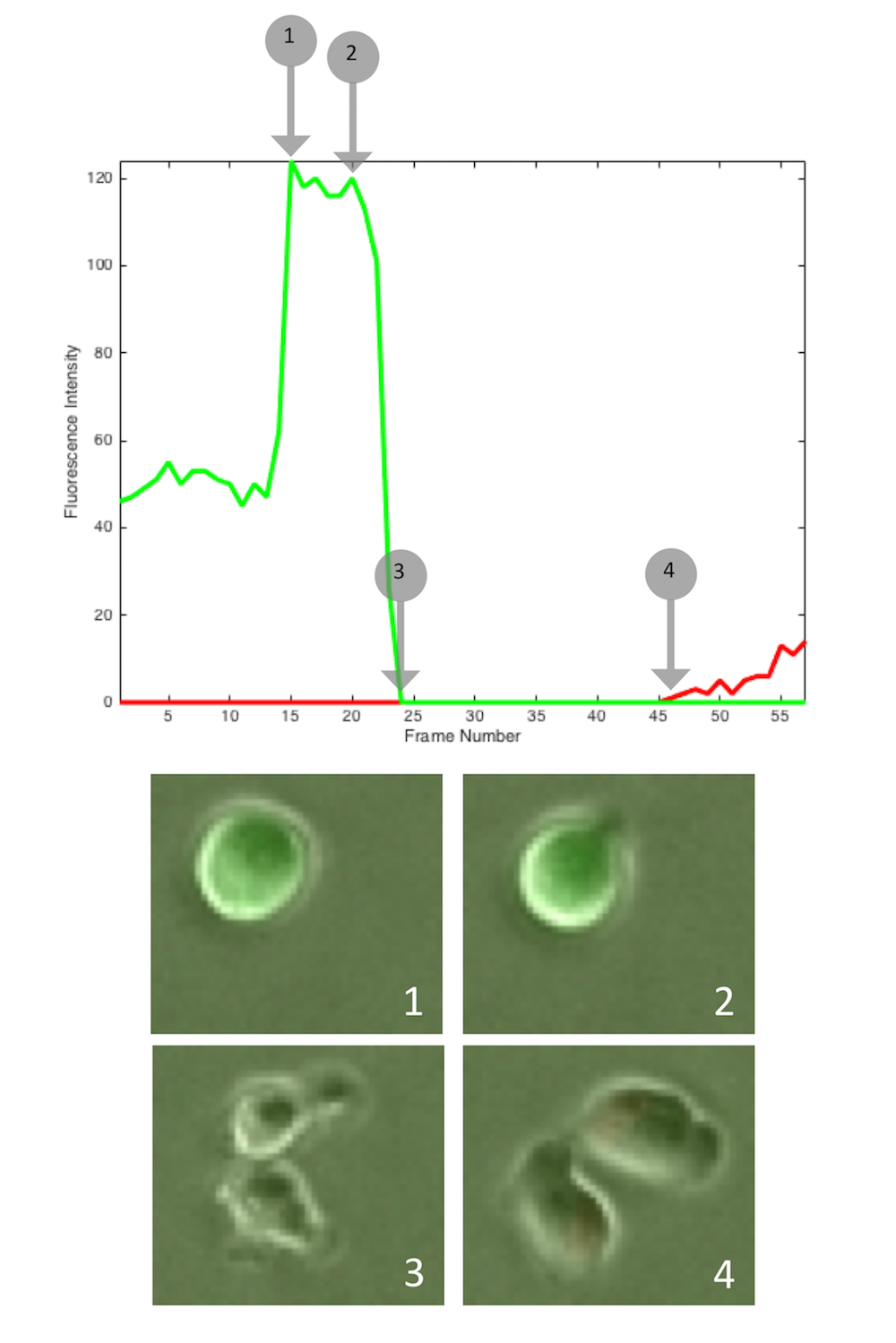}}
\subfigure[Position 3 - DMSO control]{\includegraphics[height=6.5cm]{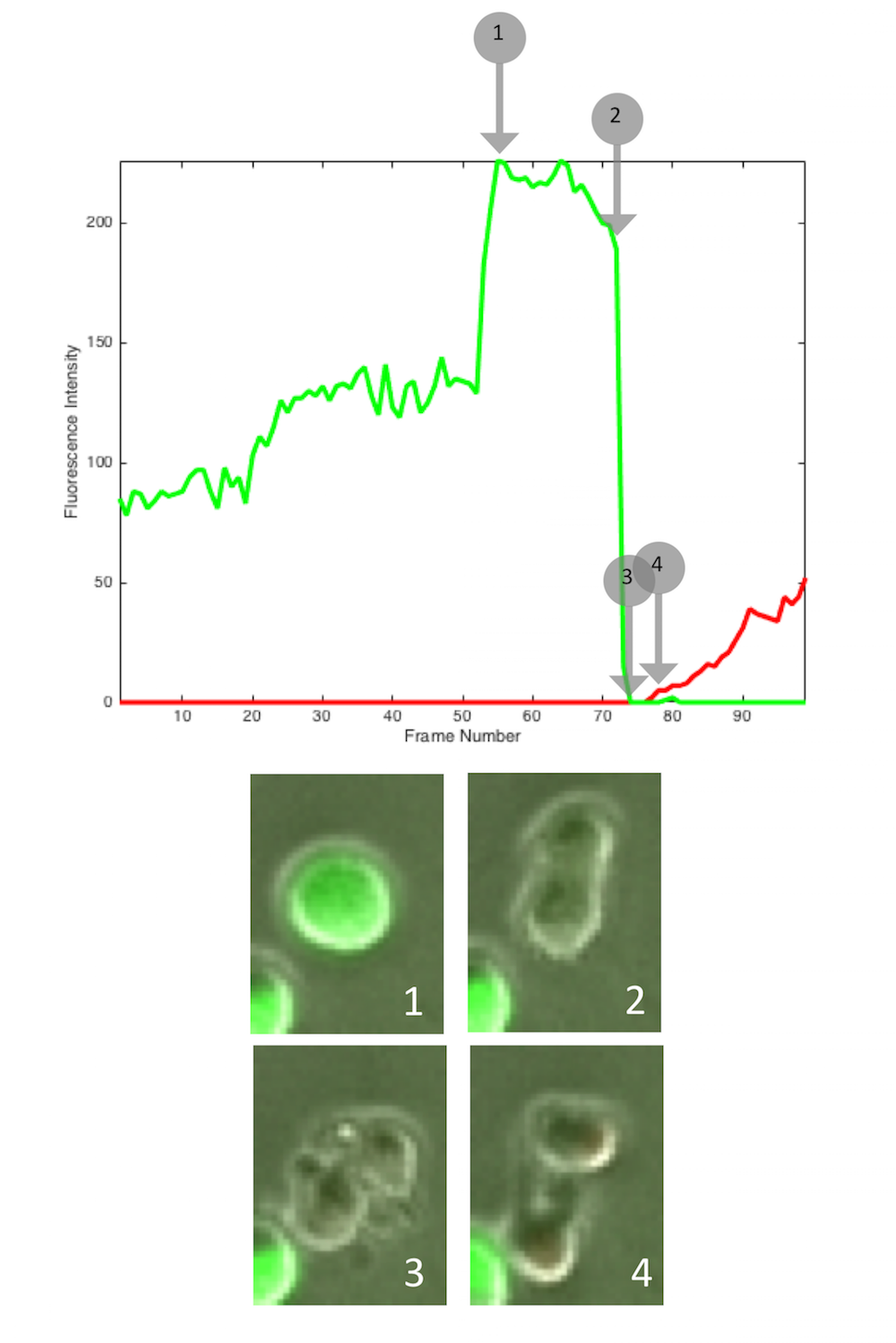}}
\label{fig:plots_fluorescence_intensities}
\end{figure}
\begin{figure}[h!]
\centering
\subfigure[Position 4 - 3nM paclitaxel]{\includegraphics[height=6.5cm]{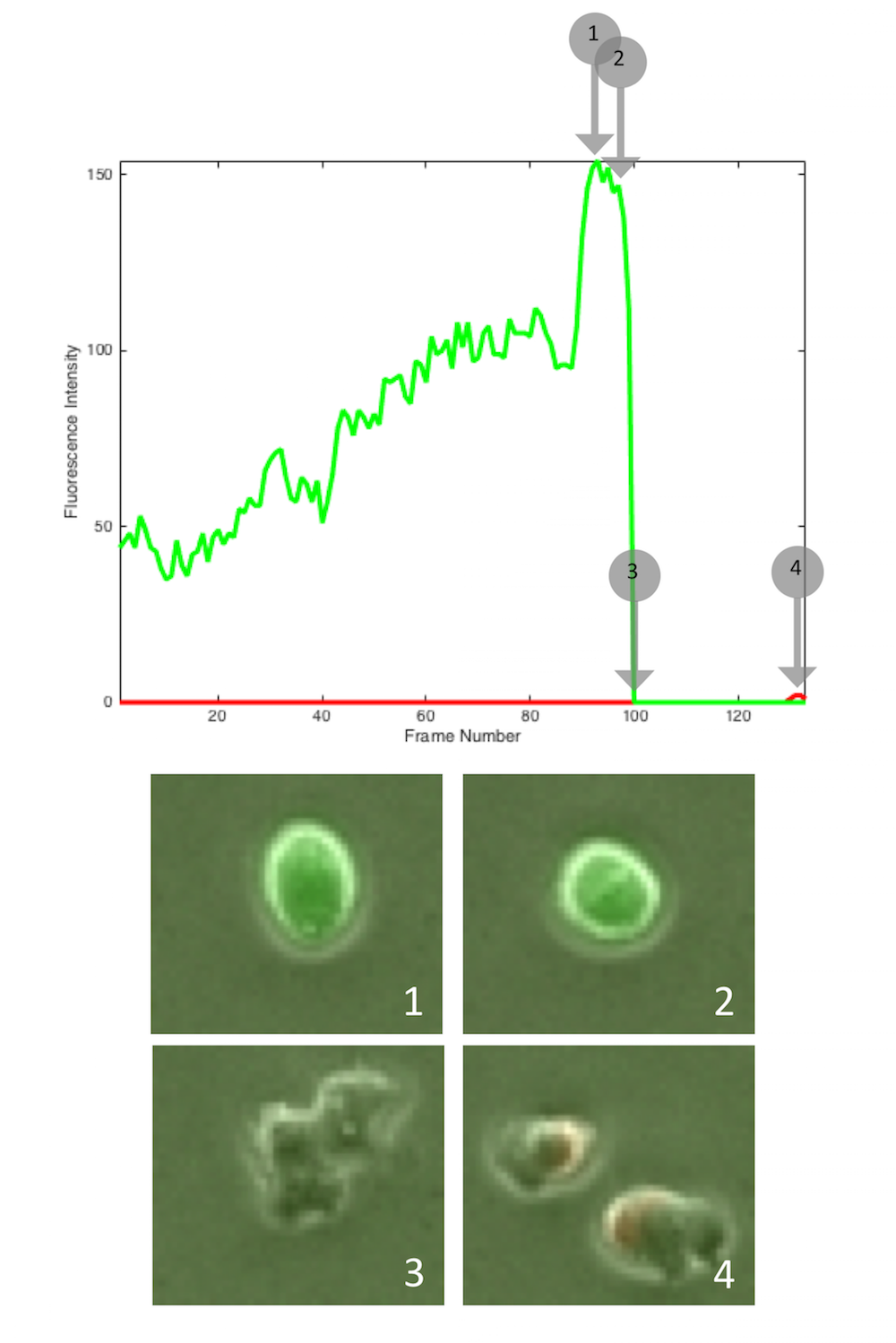}}
\subfigure[Position 5 - 3nM paclitaxel]{\includegraphics[height=6.5cm]{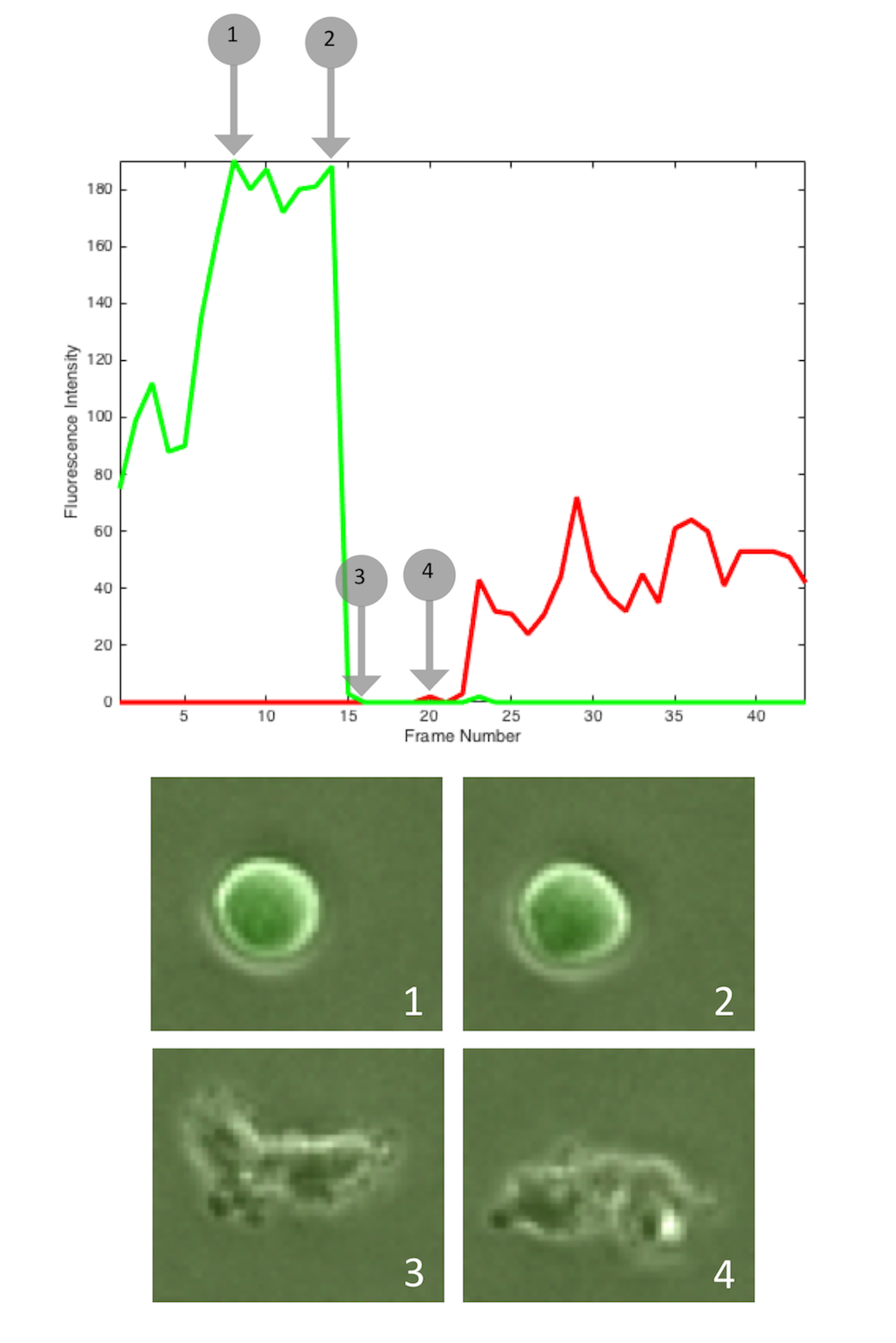}}
\subfigure[Position 6 - 3nM paclitaxel]{\includegraphics[height=6.5cm]{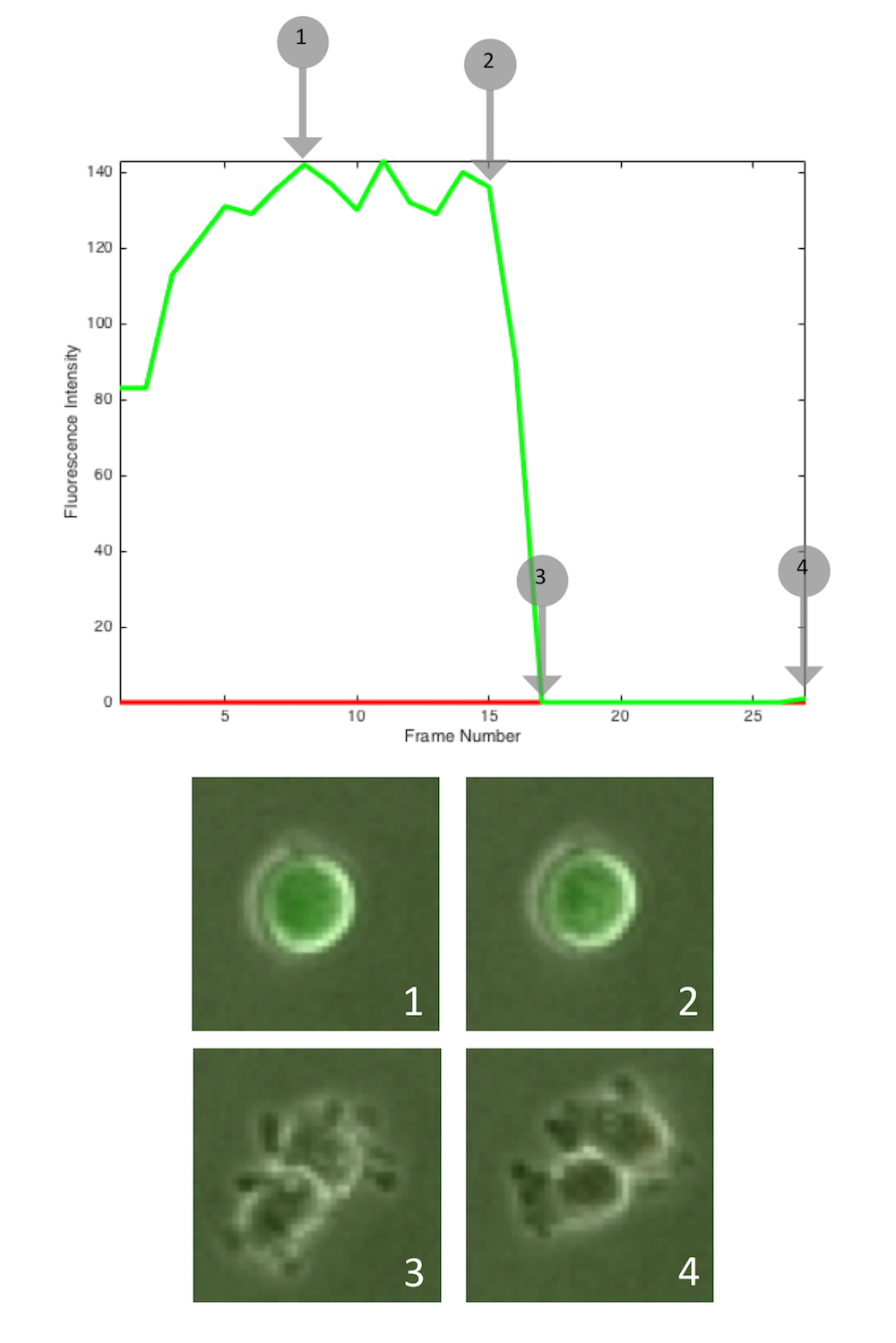}}
\\
\subfigure[Position 7 - 30nM paclitaxel]{\includegraphics[height=6.5cm]{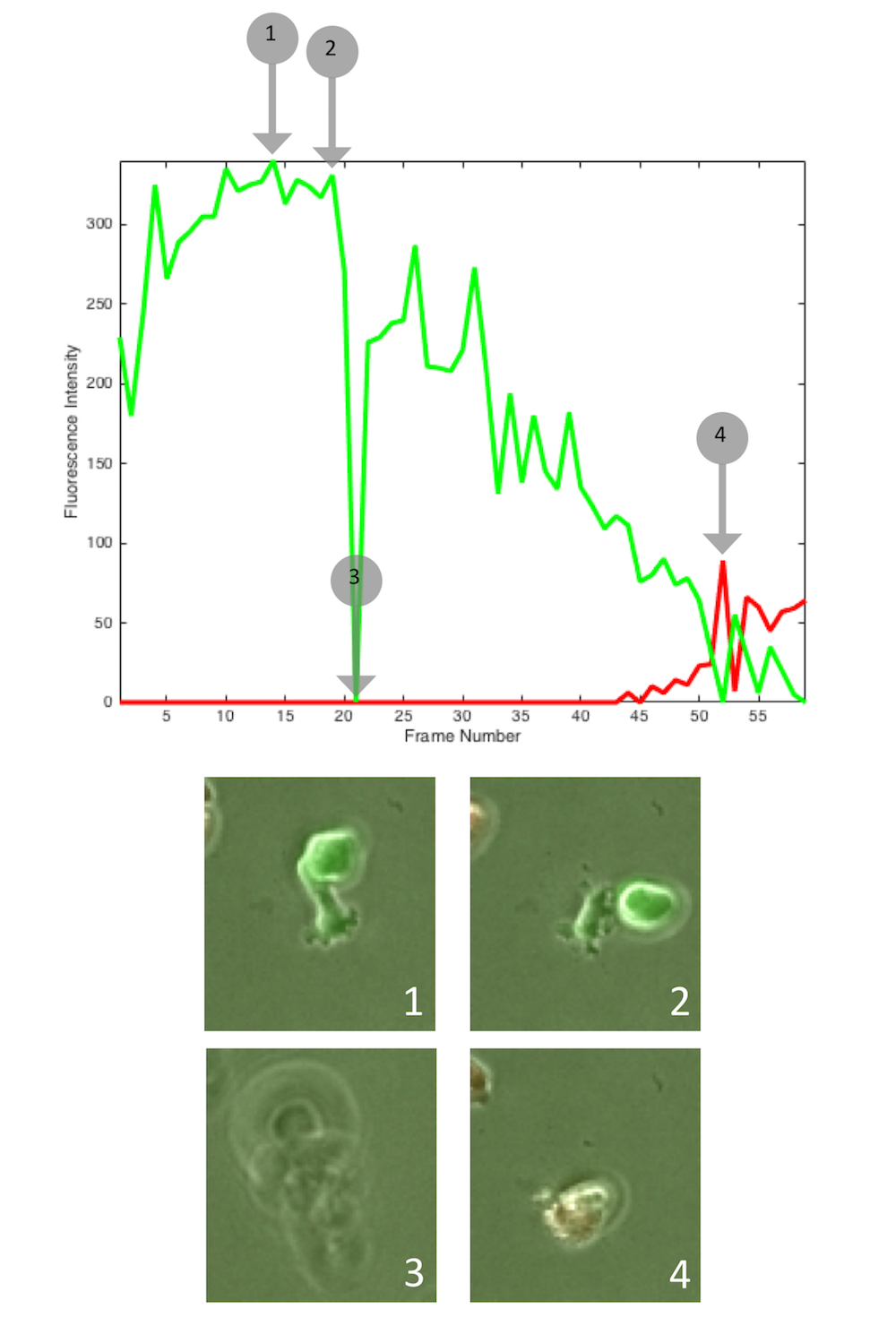}}
\subfigure[Position 8 - 30nM paclitaxel]{\includegraphics[height=6.5cm]{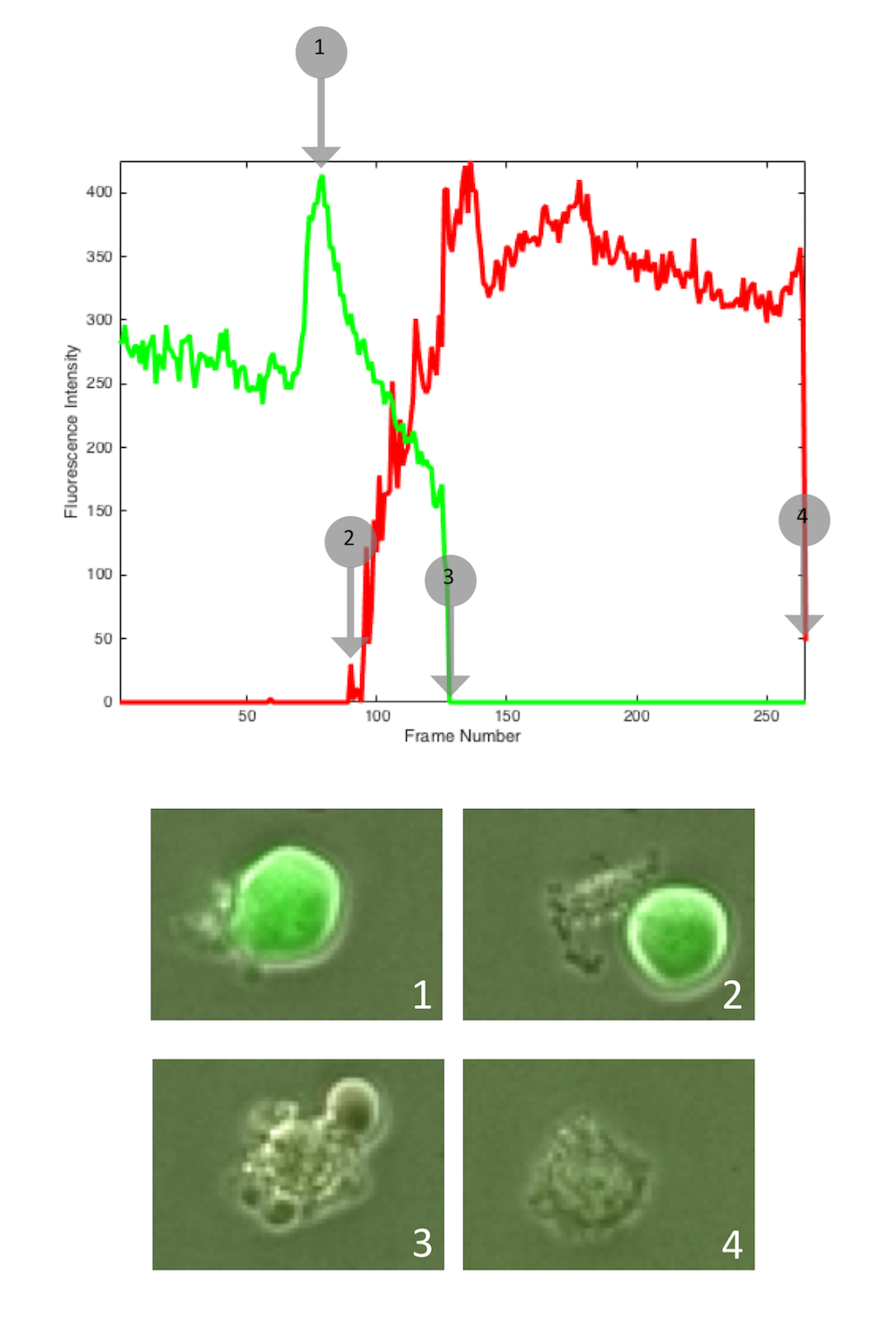}}
\subfigure[Position 9 - 30nM paclitaxel]{\includegraphics[height=6.5cm]{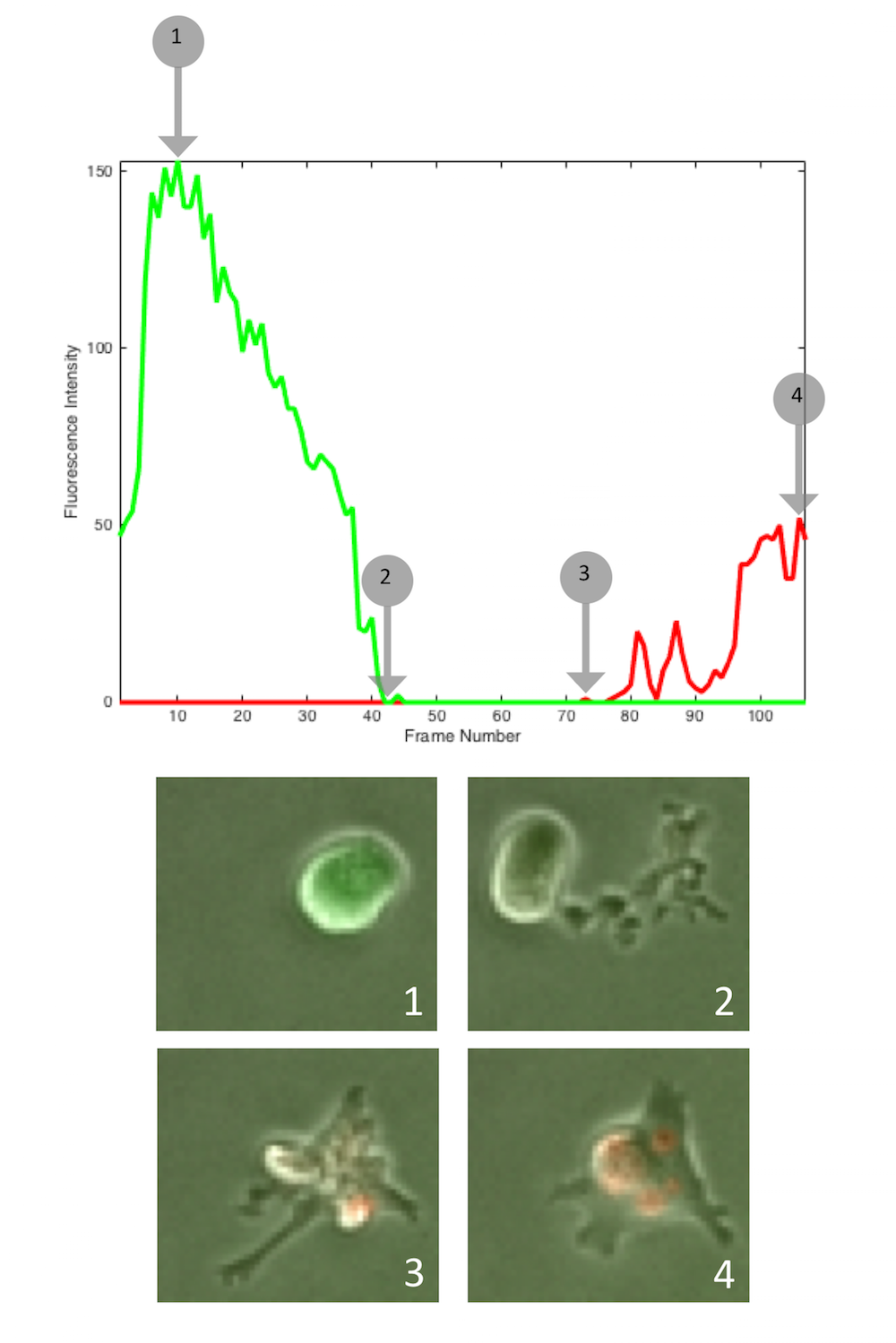}}
\caption{Fluorescence intensity distributions. We can observe the red (CDT1) and green (Geminin) fluorescence intensity distributions for nine mitotic events over time, where the cell in (h) eventually dies. To be most accurate, they were obtained by manual analysis. The peaks of green fluorescence intensity indicate mitotic cells, which are about to divide. At this point in time, where circularity is maximal as well, cells are detected as being mitotic. The backwards tracking procedure stops as soon as the green fluorescent intensity drops significantly, which can be observed in the plots on the left hand sides of positions "1". After that, the forwards tracking procedure starts again where the mitotic cell has been detected and is stopped as soon as two daughter cells have been detected by means of the circular Hough transform or the green fluorescence intensity decreased below a small threshold, i.e. becomes almost zero. The drop of green intensity in position "3" in (g) is due to an image acquisition artefact. Note that information provided by the green fluorescence intensity images is sufficient to define beginning and end of mitosis and that we do not use the red fluorescence intensity image sequences. In \emph{MitosisAnalyser}, the green fluorescence images are pre-processed by the morphological operation of erosion and thresholding.}
\label{fig:plots_fluorescence_intensities}
\end{figure}

\begin{figure}[h!]
\centering
\subfigure[Normalised parallel coordinates]{\includegraphics[height=6.5cm]{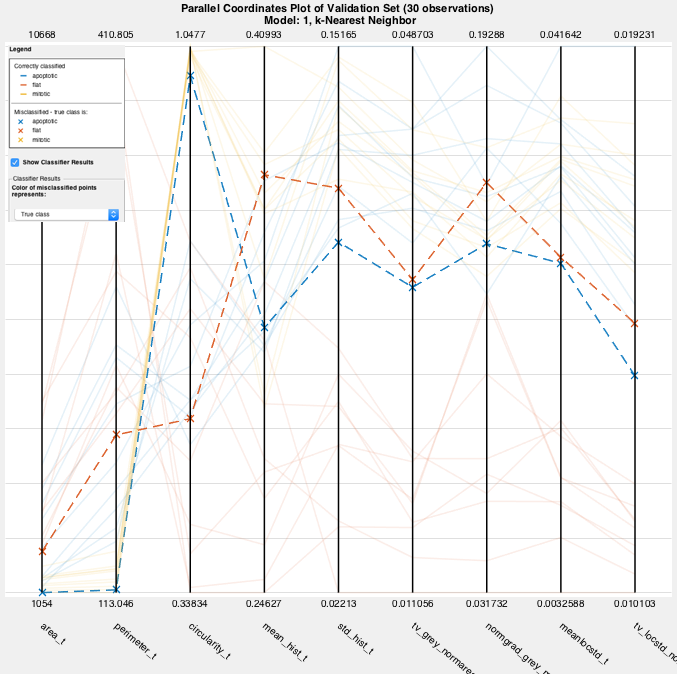}}\hspace{0.5cm}
\subfigure[Confusion matrix]{\includegraphics[height=6.5cm]{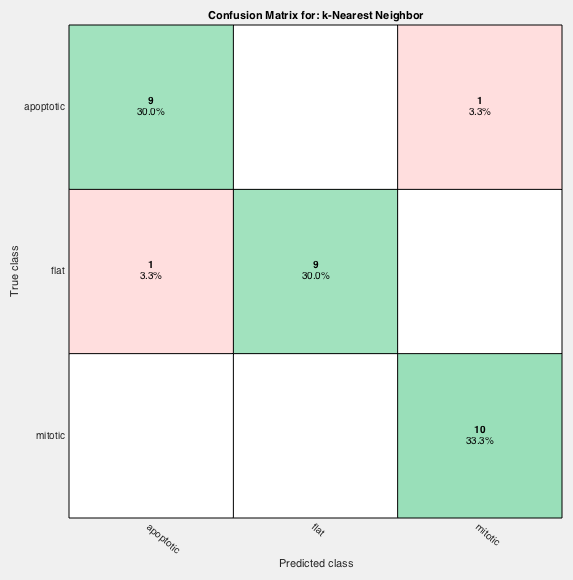}}
\caption{KNN classifier}
\label{fig:plotsClassification}
\end{figure}

\begin{figure}[h!]
\centering
\includegraphics[height=7cm]{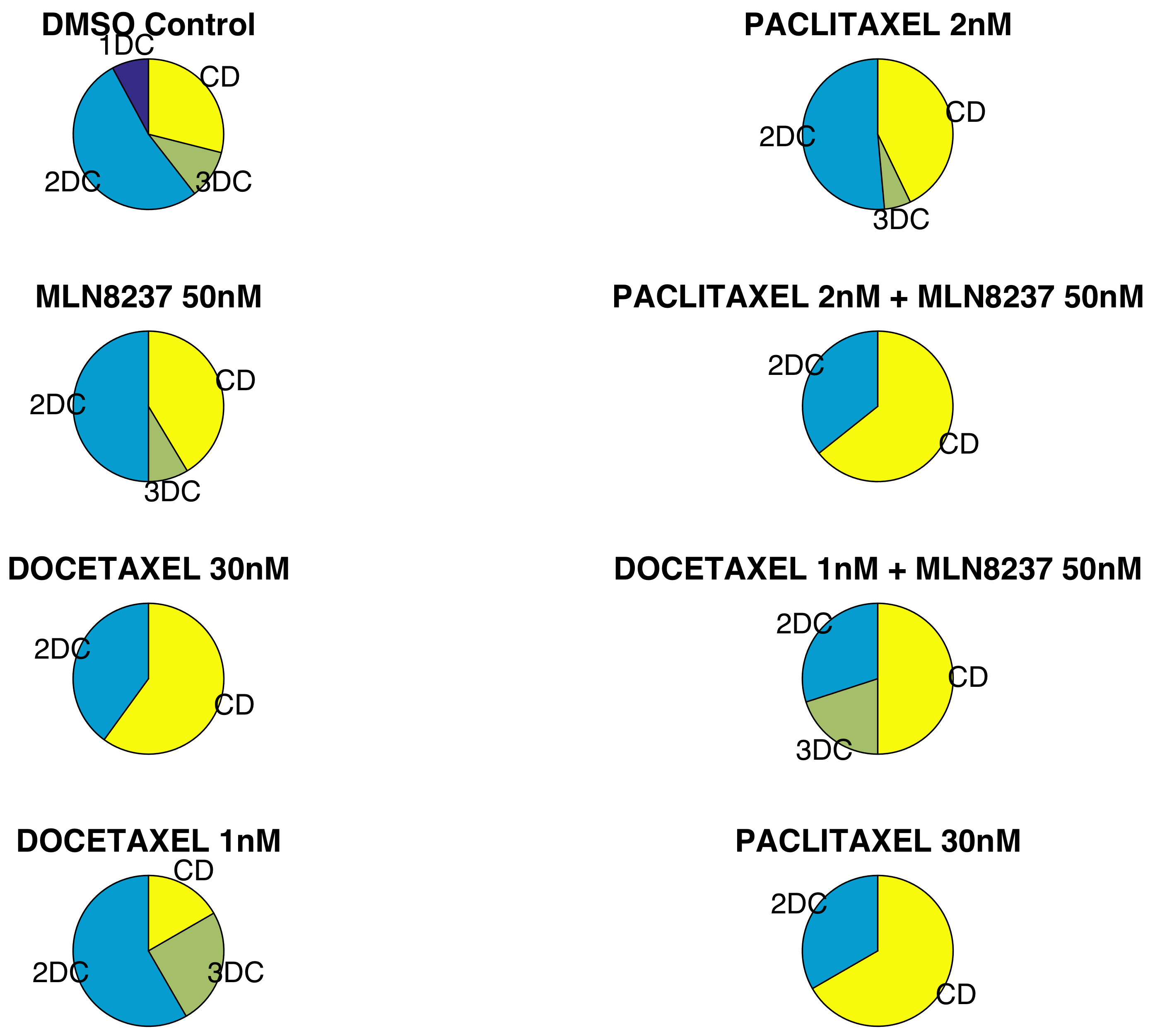}
\caption{Pie charts for fates of T24 cells treated with different drugs (1DC - 1 daughter cell, 2DC - 2 daughter cells, 3DC - 3 daughter cells, CD - cell death), where the number of analysed events is (from top left to bottom right): 38, 36, 58, 14, 5, 10, 12, 3}
\label{fig:pieChartsT24}
\end{figure}

\clearpage
\begin{table}[h!]
\centering
\begin{tabular}{|c|p{6cm}||c|c|c|}
\hline
Parameter & Description & MIA PaCa-2 & HeLa Aur A & T24\\
\hline
\hline
radiusMin & Minimum radius of mitotic cell & 10 & 10 & 10\\
\hline
radiusMax & Maximum radius of mitotic cells & 20 & 25 & 20\\
\hline
sensitivity & The higher, the more circular objects are detected & 0.8 & 0.7 & 0.7\\
\hline
mitosisThreshold & Maximum mitosis duration & 50 & 25 & 25\\
\hline
\hline
$\lambda_1$ & Weight for normal velocity term inside of cell & 1 & 0.5 & 5\\
\hline
$\lambda_2$ & Weight for normal velocity term in the background & 1 & 0.1 & 5\\
\hline
$\mu$ & Weight for length regularisation (smoothness) & 10 & 8 & 17.5\\
\hline
$\nu$ & Weight for local standard deviation term & 10 & 12 & 17.5\\
\hline
g\_adj\_low & Lower bound for rescaling of local standard deviation image & 0.08 & 0.05 & 0.08\\
\hline
g\_adj\_high & Upper bound for rescaling of local standard deviation image & 0.12 & 0.20 & 0.12\\
\hline
$\omega$ & Weight for area regularisation & 1 & 1 & 10\\
\hline
timeStep & Time step $\Delta t$ in gradient descent equation & \multicolumn{3}{|c|}{1}\\
\hline
maxIterations & Maximum number of iterations for tracking contour evolution & 5000 & 2500 & 5000\\
\hline
phiUpdate & Frequency of reinitialisation of level-set function with signed distance function & 50 & 10 & 50\\
\hline
$\varepsilon_{|\nabla|\text{Reg}}$ & Parameter in regularisation of gradient magnitude & \multicolumn{3}{|c|}{0.0001}\\
\hline
$\varepsilon_{\delta\text{Reg}}$ & Parameter in regularisation of Dirac delta function & \multicolumn{3}{|c|}{2}\\
\hline
\end{tabular}
\caption{Mitosis detection and tracking parameters for different experiments}
\label{tab:parameters}
\end{table}

\begin{table}[h]
\centering
\begin{tabular}{|c|c||c||c||c|}
\hline & &  MiA & CV & GAC\\
\hline
\hline
\multirow{2}{*}{1} & JSC & 0.6103 & \textbf{0.7056} & 0.4960\\
& MHD & 2.1713 & \textbf{0.9120} & 2.5202\\
\hline
\multirow{2}{*}{2} & JSC & 0.5736 & \textbf{0.6276} & 0.3412\\
& MHD & 3.1590 & \textbf{1.3452} & 4.8249\\
\hline
\multirow{2}{*}{3} & JSC & \textbf{0.6093} & 0.4678 & 0.3561\\
& MHD & \textbf{2.1612} & 3.0814 & 5.5271\\
\hline
\multirow{2}{*}{4} & JSC & \textbf{0.6741} & 0.4493 & 0.2848\\
& MHD & \textbf{1.7944} & 3.0597 & 6.5961\\
\hline
\multirow{2}{*}{5} & JSC & \textbf{0.4428} & 0.4243 & 0.2608\\
& MHD & 4.5354 & \textbf{2.8466} & 6.8695\\
\hline
\multirow{2}{*}{6} & JSC & \textbf{0.7133} & 0.5030 & 0.3716\\
& MHD & \textbf{0.8024} & 3.3760 & 6.0406\\
\hline
\multirow{2}{*}{7} & JSC & \textbf{0.6623} & 0.3160 & 0.4835\\
& MHD & \textbf{1.6662} & 4.3808 & 3.9340\\
\hline
\multirow{2}{*}{8} & JSC & \textbf{0.5402} & 0.4014 & 0.4541\\
& MHD & 3.5367 & \textbf{3.2165} & 4.6687\\
\hline
\multirow{2}{*}{9} & JSC & \textbf{0.5417} & 0.1597 & 0.4175\\
& MHD & \textbf{2.0813} & 6.5744 & 5.4570\\
\hline
\multirow{2}{*}{10} & JSC & \textbf{0.6877} & 0.2445 & 0.4496\\
& MHD & \textbf{0.7584} & 5.9506 & 4.7614\\
\hline
\hline
\multirow{2}{*}{avg} & JSC & \textbf{0.6050} & 0.4299 & 0.3915\\
& MHD & \textbf{2.2666} & 3.4743 & 5.1200\\
\hline
\end{tabular}
\caption{Quality measures JSC and MHD for segmentation of apoptotic cell images with \emph{MitosisAnalyser} (MiA), the model by Chan and Vese (CV) and Geodesic Active Contours (GAC) in comparison with manual segmentation}
\label{tab:qualityMeasures}
\end{table}

\begin{algorithm}[h]
\caption{Mitosis Detection}
\label{alg:md}
{
\begin{algorithmic}
\Require Image sequence, radiusMin, radiusMax, sensitivity, mitosisThreshold, distanceThreshold
	\For{every frame}
		\State Search for circularly shaped cells:
		\State [centres, radii, metrics] = \texttt{imfindcircles} (image, [radiusMin radiusMax], 'Sensitivity', sensitivity)
	\EndFor
	\State Delete circles with centre coordinates that are identical or closer than the largest detected radius
	\State If applicable, compare to circles that have already been detected in previous frames: 			\State - number of frames to be checked specified by mitosisThreshold
	\State - maximum distance of centre coordinates specified by distanceThreshold
\Ensure Centres, radii, metrics, boundaries and frame numbers of detected mitotic cells
\end{algorithmic}
}
\end{algorithm}

\begin{algorithm}[h]
\caption{Backwards Tracking}
\label{alg:bt}
{
\begin{algorithmic}
\Require Phase contrast images, [green fluorescent channel images,] initial contours, mitosisThreshold, $\lambda_1$, $\lambda_2$, $\mu$, $\nu$, $\omega$, timeStep, maxIterations, phiUpdate
\State Initialise level set function from circles surrounding mitotic cells
\While{current frame within mitosisThreshold \textbf{and} circularity $>$ threshold [\textbf{and} green intensity $>$ threshold]}
	\State Reinitialise level set function by slightly increasing previous one
	\State Create region of interest around cell interior in order to increase speed
	\State Calculate function $g$
		\While{phi does not change significantly anymore or maxIterations reached}
			\State Update phi by gradient descent:
				\State $\phi^{\text{upd}} = \phi - \Delta t \cdot \delta_{\varepsilon}(\phi) \cdot \left( \mu \nabla \cdot \left( \frac{\nabla \phi}{\vert \nabla \phi \vert} \right) + \nu \vert \nabla \phi \vert \nabla \cdot \left( g\ \frac{\nabla \phi}{\vert \nabla \phi \vert} \right) + \omega P(\phi) \right)$
				\State $P(\phi) = 
				\begin{cases}
				\int_{\Omega} \left( 1 - H(\phi(x)) \right)\,dx - t_{\text{area}},
				& \int_{\Omega} \left( 1 - H(\phi(x)) \right)\,dx \leq t_{\text{area}},\\
				0, & \text{otherwise}
				\end{cases}$
			\State Perform topology preservation combined with narrow band method
			\State Reinitialise phi every phiUpdate iterations
		\EndWhile
\EndWhile
\Ensure Contours, statistics
\end{algorithmic}
}
\end{algorithm}

\begin{algorithm}[h]
\caption{Forwards Tracking}
\label{alg:ft}
{
\begin{algorithmic}
\Require Phase contrast images, [green fluorescent channel images,] initial contours, radius, mitosisThreshold, $\lambda_1$, $\lambda_2$, $\mu$, $\nu$, $\omega$, timeStep, maxIterations, phiUpdate
\While{current frame within mitosisThreshold \textbf{and} outcome is unknown}
	\State Reinitialise level set function by slightly increasing previous one
	\State Create region of interest around cell interior in order to increase speed
	\State Calculate function $g$
		\While{phi does not change significantly anymore or maxIterations reached}
			\State Update phi by gradient descent (see Supplementary Algorithm \ref{alg:bt})
			\State Perform topology preservation combined with narrow band method
			\State Reinitialise phi every phiUpdate iterations
		\EndWhile
	\State Apply \texttt{imfindcircles} again in a region of interest around the segmented cell
	\State Determine outcome based on detected circles and statistics / classification [and green fluorescent intensities]
\EndWhile
\Ensure Contours, statistics, outcome
\end{algorithmic}
}
\end{algorithm}

\end{document}